\documentclass[aps,onecolumn,pre,showpacs,amsmath,amssymb,floatfix,showpacs,byrevtex,notitlepage,longbibliography,10pt]{revtex4-1}

\usepackage{graphicx}
\usepackage{epstopdf, epsfig}
\usepackage{amsmath}
\usepackage{amsfonts}
\usepackage{amssymb} 
\usepackage{graphicx}
\usepackage{setspace}
\usepackage{graphicx}  
\usepackage{dcolumn}   
\usepackage{bm}        
\usepackage{amsmath,amssymb}   
\usepackage{color}
\usepackage{enumitem}

\DeclareMathOperator{\e}{e}

\begin{document}
\title{Forced three-wave interactions of capillary-gravity surface waves}

\author{Annette Cazaubiel } 
\author{Florence Haudin } 
\author{Eric Falcon } 
\author{Michael Berhanu } 
\affiliation{MSC, Univ Paris Diderot, CNRS (UMR 7057), 75013 Paris, France}

\email{michael.berhanu@univ-paris-diderot.fr}
 
\date{\today}

\begin{abstract}
{Three-wave resonant interactions constitute an essential nonlinear mechanism coupling capillary surface waves. In a previous work, Haudin et al. [Phys. Rev E 93, 043110 (2016)], we have characterized experimentally the generation by this mechanism of a daughter wave, whose amplitude saturates due to the viscous dissipation. Here, we show experimentally the generation of a daughter wave verifying the resonant conditions, but not the dispersion relation.} By modeling the response of the free surface at the lowest nonlinear order, we explain this observation as a forced interaction. {The bandwidth of the linear transfer function of the free surface is indeed increased by the significant viscous dissipation.} The observation of free surface excitations not following the linear dispersion relation then becomes possible.  This forced three-wave interaction mechanism could have important consequences for wave turbulence in experimental or natural systems with non negligible dissipation.
\end{abstract}

                    
\maketitle

\section{Introduction}

Ripples propagating on a water surface are often used in introductory lectures to illustrate the physics of the waves. However, the propagation of these gravity-capillary surface waves often displays  non-linear effects when the steepness of the free surface is sufficient. Among these effects the three-wave interaction mechanism is of prime importance. As the main non-linearity is quadratic, a wave $1$ and a wave $2$ will exchange energy with a third wave $3$ to form a triad. Specifically, the case of resonant triads has been considerably investigated in the literature~\citep{McGoldrick1965,Simmons1969,Craik,Hammack1993,DrazinReidBook}, because after averaging on a long time or distance, only resonant triads may lead to a significant energy transfer between the waves~\cite{Bretherton1964}. The triad is said resonant if its three components verify the resonance conditions, \textit{i.e.} the angular frequencies $\omega_i$ and the wavenumbers $k_i$ obey simultaneously to $\omega_1\pm \omega_2 \pm \omega_3=0$ and $\mathbf{k_1} \pm \mathbf{k_2} \pm \mathbf{k_3}=0$, considering also that $\omega_i$ and $k_i$ are related by the dispersion relation $\omega_i^2=g\,k_i+\gamma/\rho\,k_i^3$ (for linear waves in the deep water regime), with $g$ the gravity acceleration on Earth, $\gamma$ the air/water surface tension and $\rho$ the {water} density. Three-wave resonant interactions are the base ingredient of the statistical study of a set of random waves interacting in a weakly non-linear regime \textit{i.e.} the wave turbulence theory~\cite{Zakharovbook,Nazarenkobook}, {which found an early application in the prediction of the turbulent spectra} of wave elevation for pure capillary waves~\cite{Zakharov1967}. Recent experimental studies with a space-time measurement~\cite{Berhanu2013,Berhanu2018} have shown that although the power-laws predicted by the wave turbulence theory are observed for capillary waves excited by gravity waves, the experimental conditions do not correspond to a weakly non-linear regime. Moreover spectra are notably populated by near-resonant three-wave interactions and the viscous dissipation of waves is significant. \\
Capillary waves occur when the surface tension is the dominant restoring force opposing to the wave propagation at the free surface and  correspond to wavenumbers $k$ larger than the inverse of the capillary length $\sqrt{(\rho\,g)/\gamma}$, which corresponds typically for water to $k \gtrsim 350$\,m$^{-1}$ and to frequencies  larger than $13\,$Hz. At these scales, the viscous dissipation is far from being negligible with a typical attenuation length of order ten times the wavelength for a frequency of about $15\,$Hz~\cite{Haudin2016}. This significant level of dissipation must be then taken into account to apply the three-wave interaction mechanism to gravity-capillary waves. To describe the experiments, a linear viscous damping term is then added as perturbation in the equations of evolution of the triad components~\cite{Craik}, despite that these equations are obtained using an inviscid model. A good agreement with the theory has been observed using this approach for the colinear degenerate case (Wilton ripple)~\cite{McGoldrick1970} and in the subharmonic generation of two daughter waves from a capillary mother wave~\cite{Henderson1987_1}. In this last example, the dissipation induces a threshold amplitude to observe the resonant triad. {In a previous work~\cite{Haudin2016}, we have generated a daughter wave by a resonant three-wave mechanism caused by the interaction of two mother waves in the capillary regime.} The amplitude of the daughter wave saturates in space due to the viscous dissipation. Nevertheless, the interaction coefficient has the order of magnitude provided by the inviscid theory.  Here, we extend this study, by demonstrating first that a daughter wave is generated in some cases where the resonance conditions are incompatible with the dispersion relation due to the chosen value of the angle between the mother waves. The daughter wave indeed verifies the resonant conditions but departs from the dispersion relation. Using a weakly non-linear model, we explain this observation as a forced interaction: the product of the two mother waves forces the free surface at the angular frequency $\omega_3=\omega_1+\omega_2$ and at the wavenumber $\mathbf{k_p}=\mathbf{k_1}+\mathbf{k_2}$ even if ${k_p} \neq k_3$, with $k_3$ satisfying the dispersion relation at $\omega=\omega_3$. The dispersion relation is indeed the free response of the free surface, which behaves similarly as a classic oscillator. In forced regime, the response is maximal in the vicinity of the eigen mode, the bandwidth being increased by the dissipation. Finally, by performing a set of experiments where the angle between the two mother waves is varied, we observe the spatial evolution of the daughter wave, that we compare with the results of the model. We report an approximate agreement due to the strong hypotheses made in the model and to the perturbing effects of reflected waves.


\section{Experimental evidence of a forced three-wave resonant interactions}
\label{Firstevidence}
\subsection{{Angular condition of resonance according to the linear dispersion relation}}
\label{Angular condition}
First, we show that two mother waves of given frequencies obeying the linear dispersion relation can interact with each other through the three-wave resonant interaction mechanism, only for specific angles between the two mother waves. Let us consider the case of two mother waves with frequencies $f_1=\omega_1/(2\pi)$ and $f_2=\omega_2/(2\pi)$. {Due to the quadratic nonlinearity of the free-surface response, the product of the mother waves generates theoretically a daughter wave at the sum frequency $f_3=f_1 + f_2$. For now, we suppose that each wave $i$ satisfies the gravity-capillary linear dispersion relation in deep water regime: 
\begin{equation}
 \omega_i^2= g\,k_i+\frac{\gamma}{\rho}\, k_i ^3
 \label{RDlin}
\end{equation} 
As the frequencies are imposed, the norms of wave vectors $\mathbf{k _i}=(2\pi)/\lambda_i$ are known by inverting numerically the relation dispersion. By definition, the triad is said to be resonant if its components satisfy simultaneously the resonance conditions}:
\begin{eqnarray}
\omega_1+\omega_2 = \omega_3 &\hspace{3cm} &  \mathbf{k_1}+\mathbf{k_2}=\mathbf{k_3}
\label{Res}
\end{eqnarray}

If these relations are verified, by defining $\alpha_{1\,2}$  the angle between the two mother wave vectors $\mathbf{k_1}$ and $\mathbf{k_2}$, then the choice of mother wave frequencies $f_1$ and $f_2$ determines completely the value of this angle, which we call the resonant angle $\alpha_{1\,2\,r}$:
\begin{equation}
\mathrm{cos}(\alpha_{1\,2\,r})=\frac{k_3^2-(k_1^2+k_2^2)}{2\, k_1.k_2 }
\label{Eqalpha}
\end{equation}

\begin{table}[th]

\centering
\setlength\extrarowheight{5pt}
\begin{tabular}{|l|c|c|c|c|c|}
 \hline
Frequency of the wave $i$ (Hz) &$k_i$ (m$^{-1})$ & $\alpha_{12\,r}$ & $v_{\phi\,i}$ (m s$^{-1}$)&  $v_{g\,i}$ (m s$^{-1}$) & $\delta_i$ (s$^{-1})$ \\
 \hline
$f_1=15$ & 428 & - & 0.220 & 0.227 & 1.47
\\
 \hline
$f_2=18$ & 507 &  - & 0.223 & 0.248 & 1.91
\\
 \hline
$f_3=f_1+f_2=33 $ & 834 & $53.8^\circ \approx 54^\circ $ & 0.249 & 0.326 & 4.25\\
 \hline
 \hline
$f_1=16$ & 455 & - & 0.221 & 0.234 & 1.61
\\
 \hline
$f_2=23$ & 626 & - & 0.231 & 0.278 & 2.66
\\
 \hline
$f_3=f_1+f_2=39$ & 774 & $58.8^\circ \approx 59^\circ$ & 0.259 & 0.349 & 5.23
\\
 \hline
\end{tabular}

\caption{{Parameters of the two triads investigated. Wave numbers $k_i$ from Eq.~(\ref{RDlin}), resonant angle $\alpha_{12\,r}$ from Eq.~(\ref{Eqalpha}), phase velocity  $v_{\phi\,i}=\omega /k$, group velocity $v_{g\,i}=\partial \omega / \partial k$, viscous damping coefficients $\delta_i=\sqrt{2\,\nu\omega_i}\,k_i/4$ (inextensible film model)~\cite{Lamb1932,VanDorn,Haudin2016}. The different values are calculated using the linear dispersion relation Eq.~(\ref{RDlin}) and the typical parameters for an air/water interface $\gamma=60$ mN/m, $\rho=1000$\,kg m$^{-3}$ and $\nu=1.00 \times 10^{-6}$ m$^2$ s$^{-1}$. The values for the actual experiments can vary of about $10$\,\% due to the variations of the fluid density with the temperature, the change of surface tension by interface contamination and the larger value of the viscosity for the intralipid solution used for space-time measurements for which $\nu=1.24 \times 10^{-6}$ m$^2$ s$^{-1}$~\cite{Haudin2016}. The values of the resonant angles are then approximated in the rest of this article $\alpha_{12\,r}=54^\circ$ for the first triad and $\alpha_{12\,r}=59^\circ$ for the second.}}
\label{Tab1}
\end{table}

\begin{figure}
\begin{center}

\includegraphics[height=0.43\columnwidth]{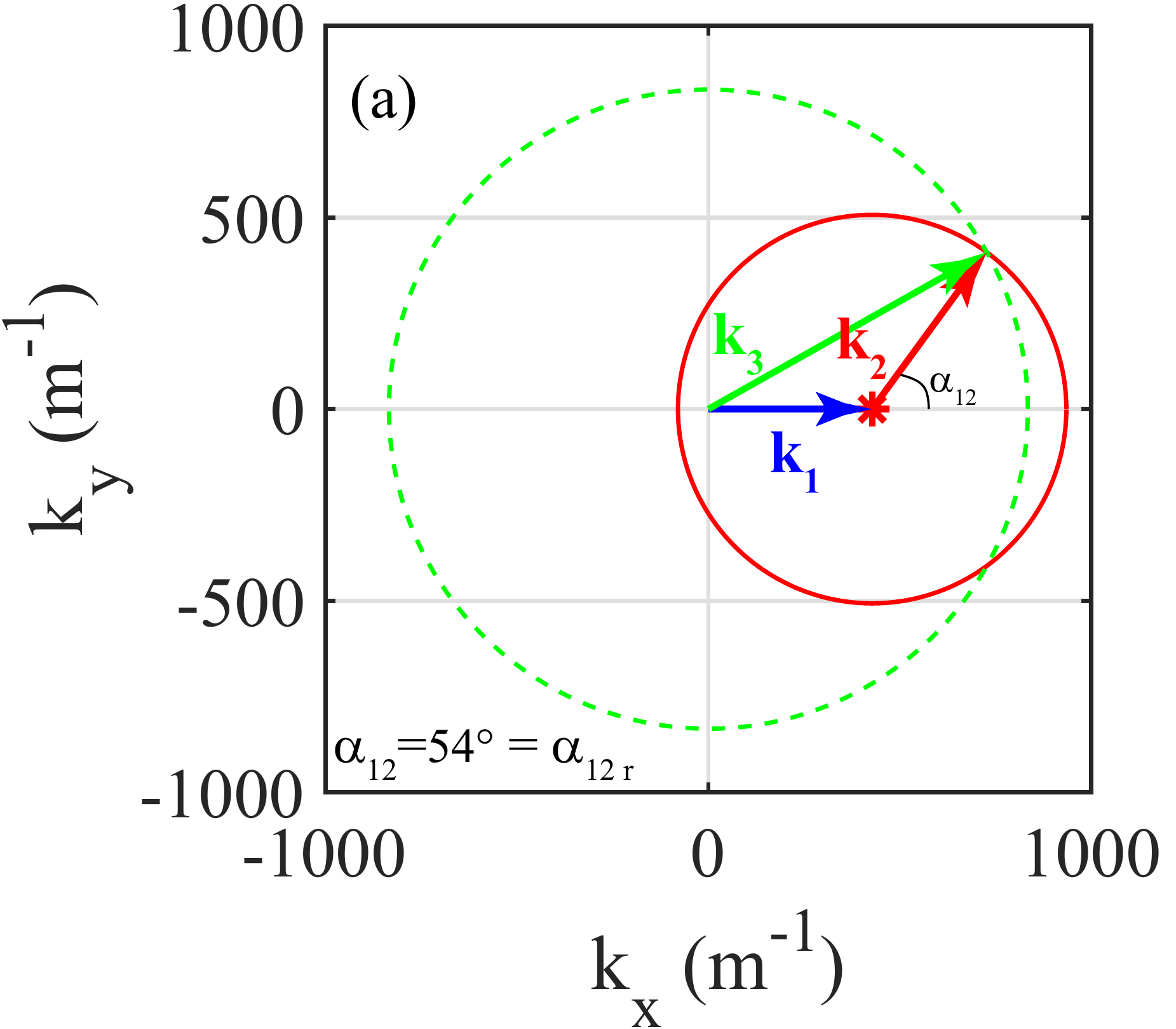}\hfill
\includegraphics[height=0.43\columnwidth]{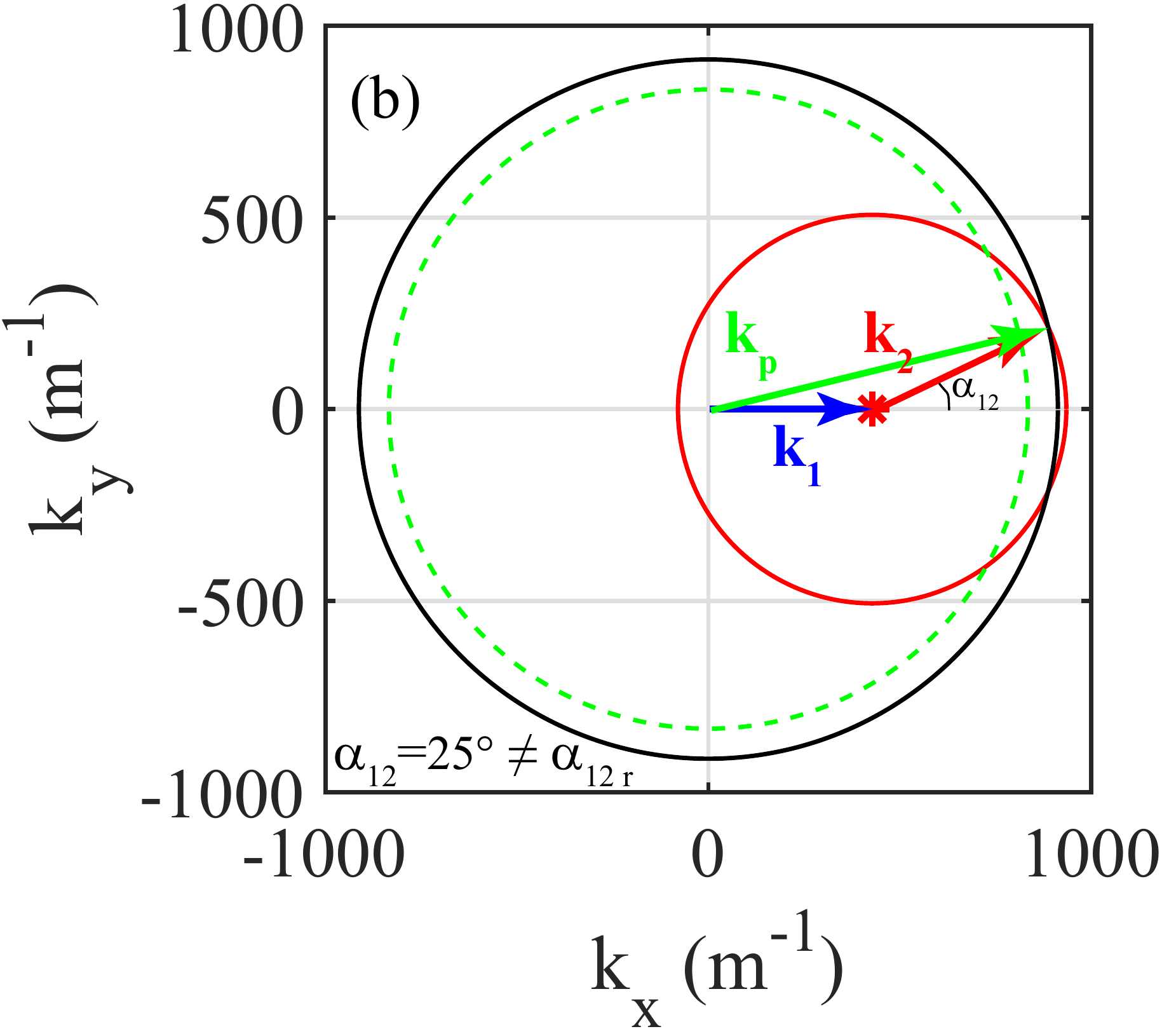}
\caption{Triad for $f_1=15$ Hz and $f_2=18$ Hz with (a) $\alpha_{1\,2\,r}=54$ $^\circ$ (classic three-wave resonant interaction case) and (b)  $\alpha_{1\,2}=25$ $^\circ$  (forced three-wave resonant interaction case). The green circle (dashed {line}) corresponds to $\mathbf{k_3}$ satisfying the gravity-capillary dispersion relation, the red circle (light gray continuous line) to all locations of the extremity of vector $\mathbf{k_{2}}$ and the black circle (continuous line (b) only) to the modulus of the vector $\mathbf{k_{p}=\mathbf{k_1}+\mathbf{k_2}}$ when the angle between $\mathbf{k_1}$ and $\mathbf{k_2}$ is  $\alpha_{1\,2}=25$ $^\circ$. When $\alpha_{1\,2} \neq \alpha_{1\,2\,r}$, the green circle (dashed line) for the dispersion relation at $\omega_3$ and the black circle for the spatial resonance condition are distinct. The triad $\mathbf{k_1}$, $\mathbf{k_2}$ and $\mathbf{k_p}$ satisfies the resonance conditions but not the dispersion relation.}\label{triad_graphs1}
\end{center}
\end{figure}

We consider the triad investigated in our previous work~\cite{Haudin2016}, with $f_1=15\,$Hz, $f_2=18\,$Hz and $f_3=f_1+f_2=33\,$Hz, $\alpha_{1\,2\,r} \approx 54^\circ$ by taking $\gamma=60$ mN/m and $\rho=1000$ kg/m$^3$ for water as working fluid.  {For the second triad investigated $f_1=16\,$Hz, $f_2=23\,$Hz and $f_3=f_1+f_2=39\,$Hz, we have $\alpha_{1\,2\,r} \approx 59^\circ$. The parameters of both triads are indicated in Table~\ref{Tab1}}. It is also possible to determine graphically the resonant wave vector $\mathbf{k_{3r}}$ satisfying both the dispersion relation and the resonance conditions as illustrated in Fig.~\ref{triad_graphs1} (a). The red circle (continuous line) defines the loci of all the possible $\mathbf{k_3}$ built by the sum of $\mathbf{k_1}+\mathbf{k_2}$, when changing the orientation of the vector $\mathbf{k_2}$ on this circle, keeping its norm $k_2$ constant. The green circle (dash-dotted line) corresponds to the loci of the vectors $\mathbf{k_{3}}$ whose norm is given by the dispersion relation at $f_3$. As a consequence, the intersections between the red and the green circles define the vectors $\mathbf{k_{3}}$ satisfying both the resonance conditions and the relation dispersion. Two solutions exist corresponding to opposite values of $\alpha_{1\,2\,r}$. If the angle between $\mathbf{k_1}$ and $\mathbf{k_2}$ differs from $\alpha_{1\,2\,r}$, being for example $25^\circ$ ( Fig.~\ref{triad_graphs1} (b) ), then the modulus of the vector defined as $\mathbf{k_{p}=\mathbf{k_1}+\mathbf{k_2}}$ is different from $k_3$ and does not belong to the dispersion relation. Therefore the corresponding triad violates the dispersion relation. In these conditions the generation of a daughter wave is unexpected. However, for capillary-gravity waves, experimentally, we show in the next part that we observe actually the daughter wave at the frequency $f_3=f_1+f_2$, whatever the value of $\alpha_{1\,2}$. We demonstrate in the following that the generation of the daughter wave in these conditions corresponds to a forced three-wave interaction mechanism.\\

\subsection{{Experimental methods}}
\label{ExpMethods}
\begin{figure}
 \begin{center}
\begin{minipage}[c]{0.34\textwidth}
 \includegraphics[width=\textwidth]{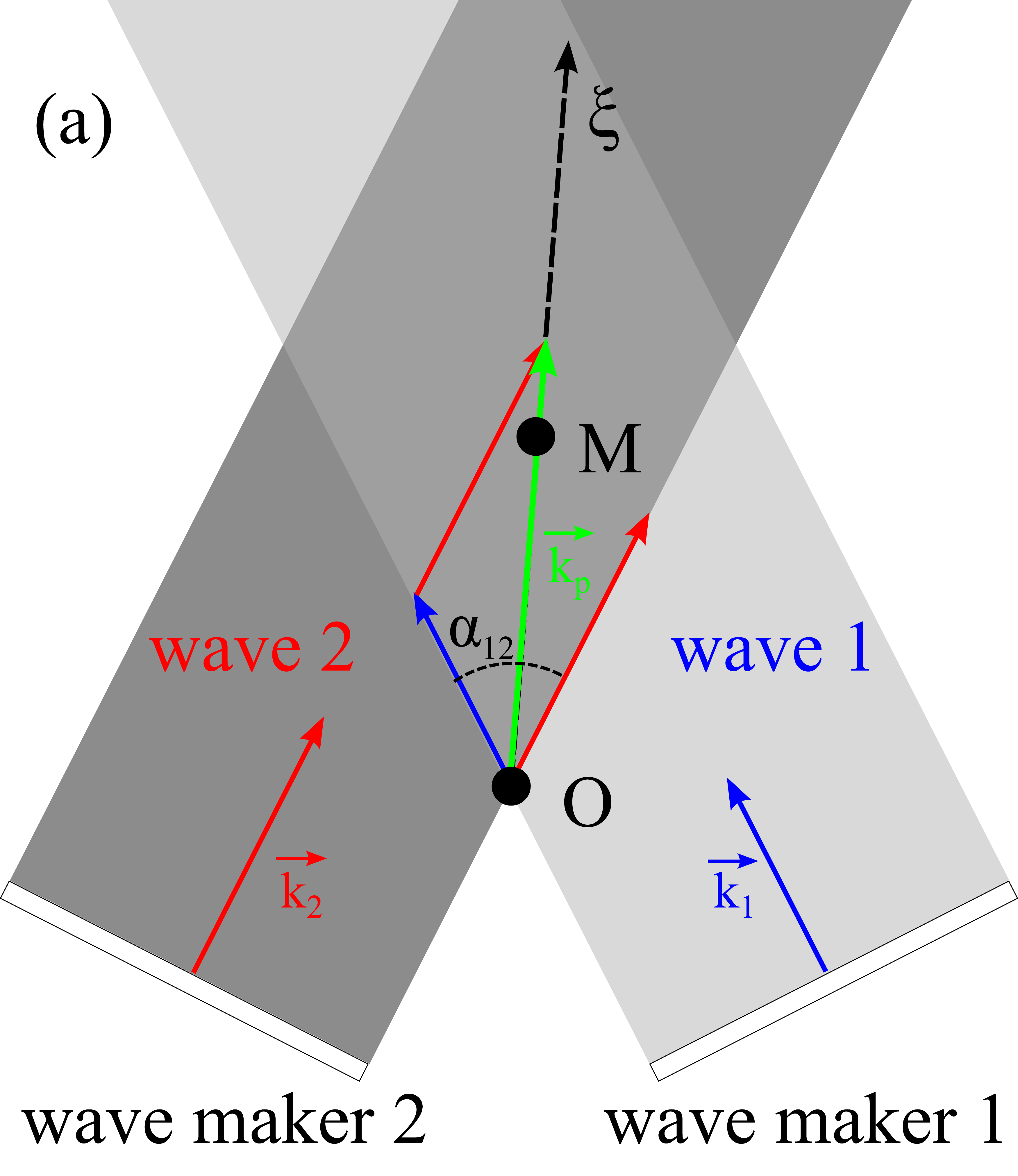}  \hspace{1cm}
  \end{minipage} \hfill 
\begin{minipage}[c]{0.48\textwidth}
  \includegraphics[width=\textwidth]{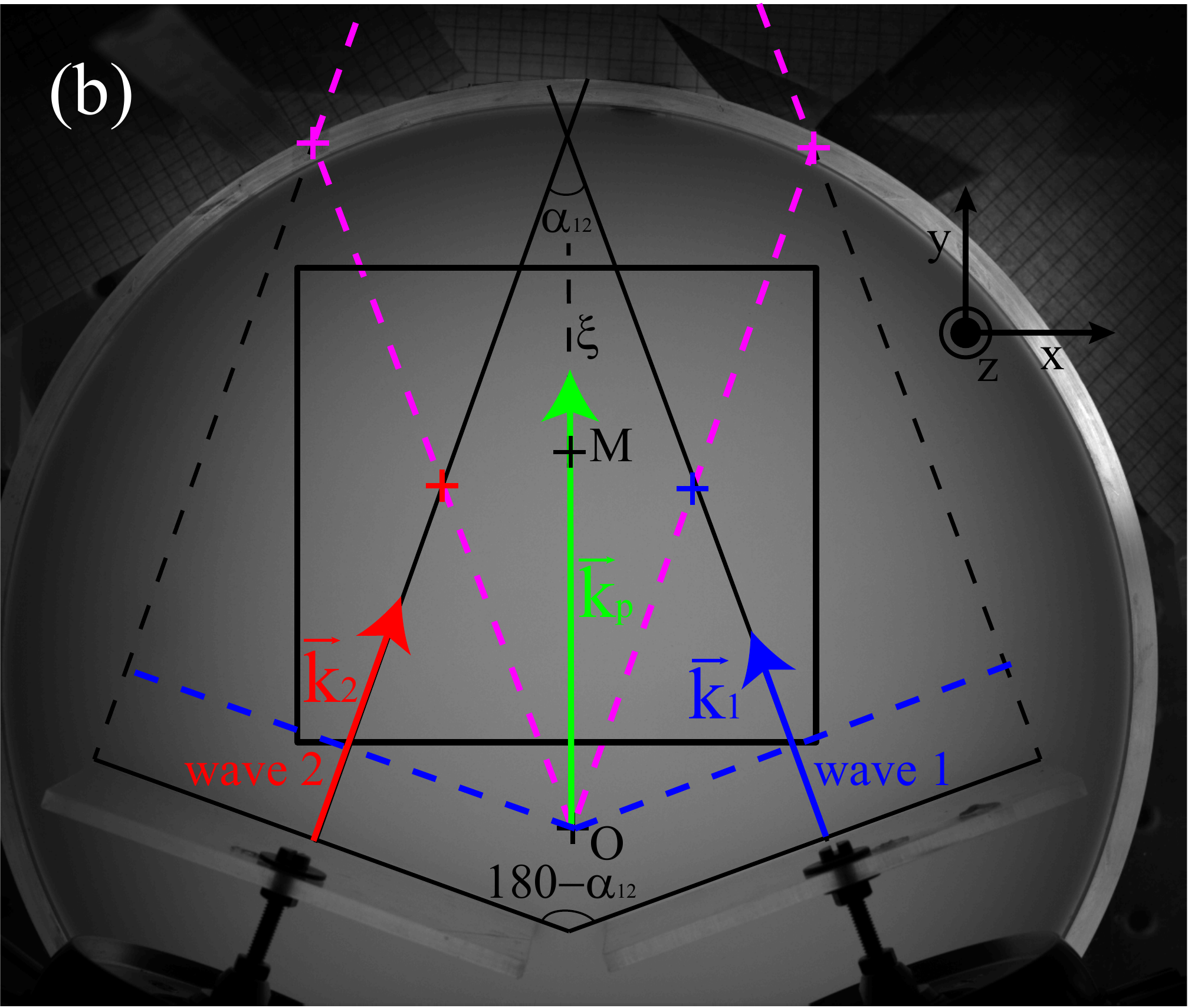}  
   \end{minipage}
 \end{center}
 \caption{(a) Schematic view of the interaction zone between the two mother waves $1$ and $2$. O is the origin of this zone and M locates a point in this area along the direction $O_\xi$ given by $\mathbf{k_p}=\mathbf{k_1} + \mathbf{k_2}$. The two mother waves cross with an angle $\alpha_{1\,2}$. (b) Top view with the CMOS fast camera of the setup for spatiotemporal measurements with the Diffusing Light Photography (DLP) method for {$\alpha_{1\,2}=40.2^\circ \approx 40^\circ$}. The liquid is a solution of Intralipids 20\% \texttrademark \, in water. The black rectangle depicts the observation area $\mathcal{S}$ of $110.5 \times 101.0$ mm$^2$ where the wave field reconstruction is performed. The magenta polygon (dashed line) delimits the interaction zone where the two mother waves $1$ and $2$ cross each other and interact. {The blue dashed lines are the parallels to the mother wave fronts passing by O.} The direction of $\mathbf{k_1}$, $\mathbf{k_2}$ and $\mathbf{k_p}$ are indicated. {$\xi$ is the distance from the point O on the line starting from O and directed by $\mathbf{k_p}$. M is a measurement point on this line located at $\xi=80\,$mm on this image. For the local measurements with the laser vibrometer $\xi \approx 70\,$mm.} The fluid is at rest in this picture.}
 \label{schema3ondes}
\end{figure}

This observation is obtained by producing in a cylindrical tank filled with water, the two mother waves of frequencies $f_1$ and $f_2$ crossing each other with an angle $\alpha_{1\,2}$. Due to the finite width of the wavemakers, the intersection of the two wave beams defines a zone (see Fig.~\ref{schema3ondes} (a)), in which we detect and characterize the daughter wave $3$.
The experimental device is identical to the one used in our previous study for $\alpha_{1\,2}=\alpha_{1\,2\,r}$~\cite{Haudin2016}. We refer to this work, for the schema of the experimental device and for more information on the experimental methods, that we recall here briefly. The two mother waves $1$ $(\mathbf{k_1},f_1)$ and $2$ $(\mathbf{k_2},f_2)$  are generated using two paddles of width $100\,$mm inside a cylindrical tank of internal diameter $240\,$mm and filled with filtered water up to a height of $50\,$mm. Each paddle is driven by an electromagnetic shaker. The deformation of the surface is studied either by using a local measurement technique (laser vibrometer Polytech OPV 506) or by performing a spatial reconstruction of the wave field, the Diffusing Light Photography (DLP) method ~\cite{Putterman1996,Putterman1997,Berhanu2013,Haudin2016}. 
For the local measurements with the vibrometer, the water is dyed in white thanks to TiO$_2$ pigment (Kronos 1001, 10 g in 1 L). Similarly, the DLP method requires adding $5$\,mL per L of Intralipid 20 \% Fresenius Kabi \texttrademark \, to make the liquid diffusive for light propagation. Then by lighting from below, the transmitted light measured by focusing a camera (PCO Edge, scientific CMOS) on the top surface provides after calibration the local depth $h(x,y)$ of the liquid. 
Each run is recorded during $40.96$\,s with a frame frequency of $200\,$Hz in stationary regime in a spatial window $\mathcal{S}=110.5 \times 101.0$\,mm$^2$ corresponding to $1120 \times 1024$ pixels (see Fig.~\ref{schema3ondes} (b)). The orientations of the wavemaker paddles are varied between the experiments and the angle between the mother waves $\alpha_{1\,2}$ is deduced from the measurement of the angle between the wavemaker paddles, which are imaged by the camera. The relative position of the interaction zone differs thus between the measurements. When the angular dependency is tested, the amplitudes of the two mother waves are kept at a constant level ($\sim 0.1$\,mm) for all measurements, sufficient to observe the generation of a daughter wave. After a static calibration, using the DLP method, the deformation of the free surface is obtained for each images $\eta (x,y,t)=h(x,y,t)-h_0$, with $h_0$ the depth of fluid at rest. The 2D spatio-temporal spectrum of wave elevation $S_\eta (k_x,k_y,\omega)$ is computed by performing a 2D spatial Fourier transform and a temporal Fourier transform on the deformation of the free surface $\eta(x,y,t)$. {Using the spatio-temporal spectrum $S_\eta(\omega,k)$ the experimental dispersion relation can be measured and we note that the value of the surface tension for the solution of intralipids is lower $\gamma = 55$\,mN.m$^{-1}$ compared to the case of pure water. The kinematic viscosity of the intralipid solution has been previously measured $\nu=1.24 \times 10^{-6}$ m$^2$ s$^{-1}$~\cite{Haudin2016}. For the local measurement with the laser vibrometer using water with TiO$_2$ pigment, the properties are closer to those of pure water~\cite{Przadka2012} and we have found $\gamma = 62$\,mN.m$^{-1}$  and $\nu=1.02 \times 10^{-6}$ m$^2$ s$^{-1}$~\cite{Haudin2016}. }

\begin{figure}
 \begin{center}
\includegraphics[height=.38\textwidth]{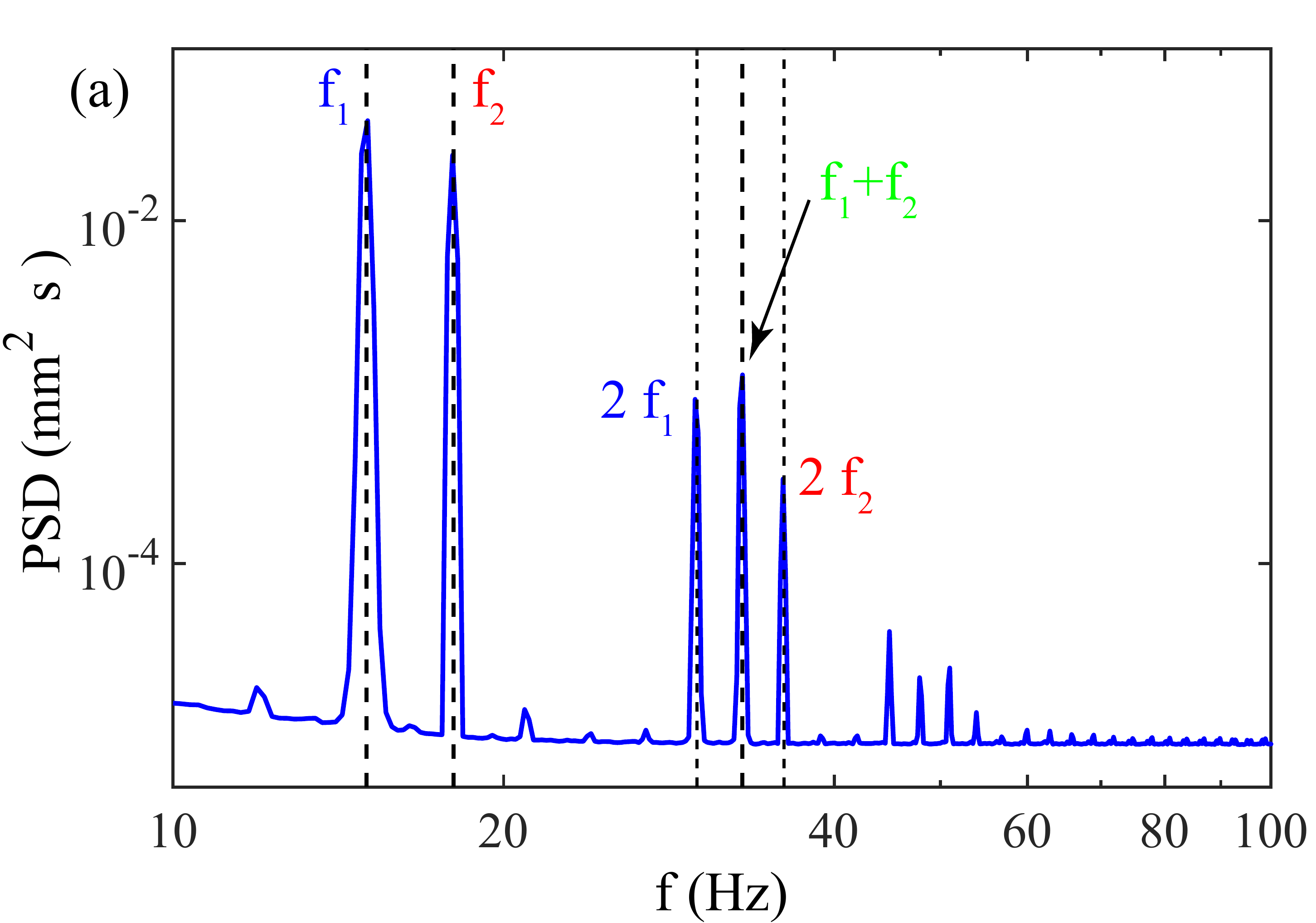}\hfill
\includegraphics[height=.39\textwidth]{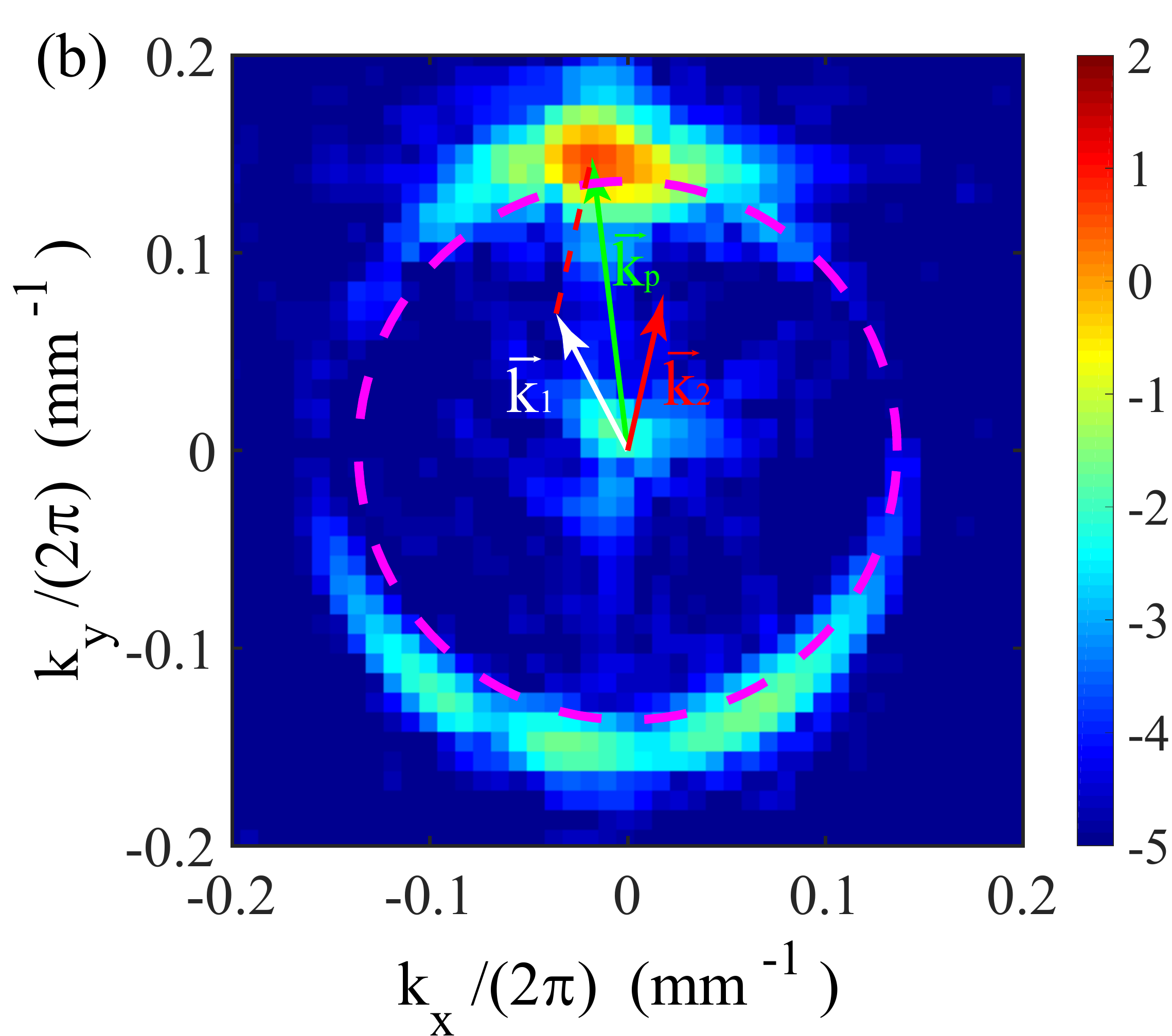}
 \end{center}
 \caption{(a) Temporal power spectrum of wave elevation $P_\eta(\omega=2\pi\,f)$ for $f_1=15$\,Hz, $f_2=18\,$Hz and $\alpha_{1\,2}=40\,^\circ$ obtained with the DLP method (spatial average on the window of observation). We observe significant peaks mainly at the forcing frequencies $f_1$ and $f_2$, at the harmonics of the forcing $2f_1$ and $2 f_2$ and at the sum frequency $f_1+f_2$ resulting from a three-wave interaction. (b) Spatio-temporal spectrum $S_\eta (k_x,k_y,\omega_3)$ plotted at the daughter wave frequency $f_3$.  The colorscale corresponds to $\log_{10} [S_\eta]$. {The arrows indicate the vector $\mathbf{k_{i\,exp}}$ extracted from the maxima of the experimental spectra $S_\eta (k_x,k_y,\omega_1)$ and $S_\eta (k_x,k_y,\omega_2)$~\cite{Haudin2016,Deike2017}. We observe $\mathbf{k_{3\,exp}} \approx \mathbf{k_{1\,exp}}+\mathbf{k_{2\,exp}} \approx \mathbf{k_p}$, but ${k_p}$ is larger than the value given by the linear dispersion relation (dash magenta circle).}}
 \label{spMES8}
\end{figure}

\subsection{{Experimental evidence of generation of a resonant daughter wave when $\alpha_{1\,2} \neq \alpha_{1\,2\,r}$}}
 \label{Firstexp}
Here, we present the results of a first set of experiments for  the triad  $(f_1=15\,\mathrm{Hz},f_2=18\,\mathrm{Hz},f_3=f_1+f_2)$ in the case $\alpha_{1\,2}=40\,^\circ$. The temporal spectrum $P_\eta(\omega)$ is defined as $P_\eta(\omega)=\int\,S_\eta (k,\omega)\,\mathrm{d}k$. $P_\eta(\omega)$ is displayed in Fig.~\ref{spMES8} (a). This graph reveals a peak at frequency $f_3=f_1+f_2$ attesting for the existence of a daughter wave. Its relative amplitude is found here {similar to the resonant case $\alpha_{1\,2}=54$ $^\circ=\alpha_{1\,2\,r}$~\cite{Haudin2016}} and also {slightly larger} than the $2f_i$ harmonics of mother waves. Using a space-time reconstruction of the wave field with the DLP method, the directions of the components of the triad can be measured~\cite{Haudin2016}, as it can be seen in Fig.~\ref{spMES8} (b) displaying a spatio-temporal spectrum for $f_3$, $S_\eta (k_x,k_y,\omega_3)$. The arrows correspond to the positions of the maximum of each spectrum, defining thus the experimental values of {$\mathbf{k_{1\,exp}}$, $\mathbf{k_{2\,exp}}$ and $\mathbf{k_{3\,exp}}$. We verify then $\mathbf{k_{3\,exp}} \approx \mathbf{k_{1\,exp}}+\mathbf{k_{2\,exp}}$. The finite size of the images implies a finite resolution in $k$ in the spatial spectra of about $84$ m$^{-1}$. The measurement uncertainty in the determination of $\mathbf{k_{i\,exp}}$
is estimated to be the half value $\pm\, 42$ m$^{-1}$. Given this uncertainty, we assume then that $\mathbf{k_{3\,exp}} \approx\mathbf{k_p}=\mathbf{k_1}+\mathbf{k_2}$ verifying the wave-number resonant condition. As the linear dispersion relation and the resonance conditions cannot be satisfied simultaneously, $\mathbf{k_{3\,exp}} > k_{3}$ ($k_3$ is given by the dispersion relation Eq.~(\ref{RDlin}) at the frequency $f_3$). The peak corresponding to $\mathbf{k_{3\,exp}}$ in Fig.~\ref{spMES8} (b)  is thus outside of the linear dispersion relation at $f_3$ represented by the magenta dashed circle {of radius $k_3$}, as ${k_ {p}} \neq k_3 $.} We observe also in our experiments that a significant signal energy is found at the wavenumber $k=k_p$, but in the opposite direction of $\mathbf{k_ {p}}$ due to the reflections of the daughter wave on the tank wall. {The same method has been used to verify the three-wave resonance spatial condition using different techniques of wave field reconstruction like the Free-Surface Synthetic Schlieren~\cite{Moisy2009} for capillary waves~\cite{Soriano2018} and the Fourier Transform Profilometry~\cite{Cobelli2009} for hydro-elastic waves~\cite{Deike2017}, but with the same limited resolution in wavenumber due to the finite size of images.}  \\
 \begin{figure}[!h]
 \begin{center}
  \includegraphics[width=0.48\columnwidth]{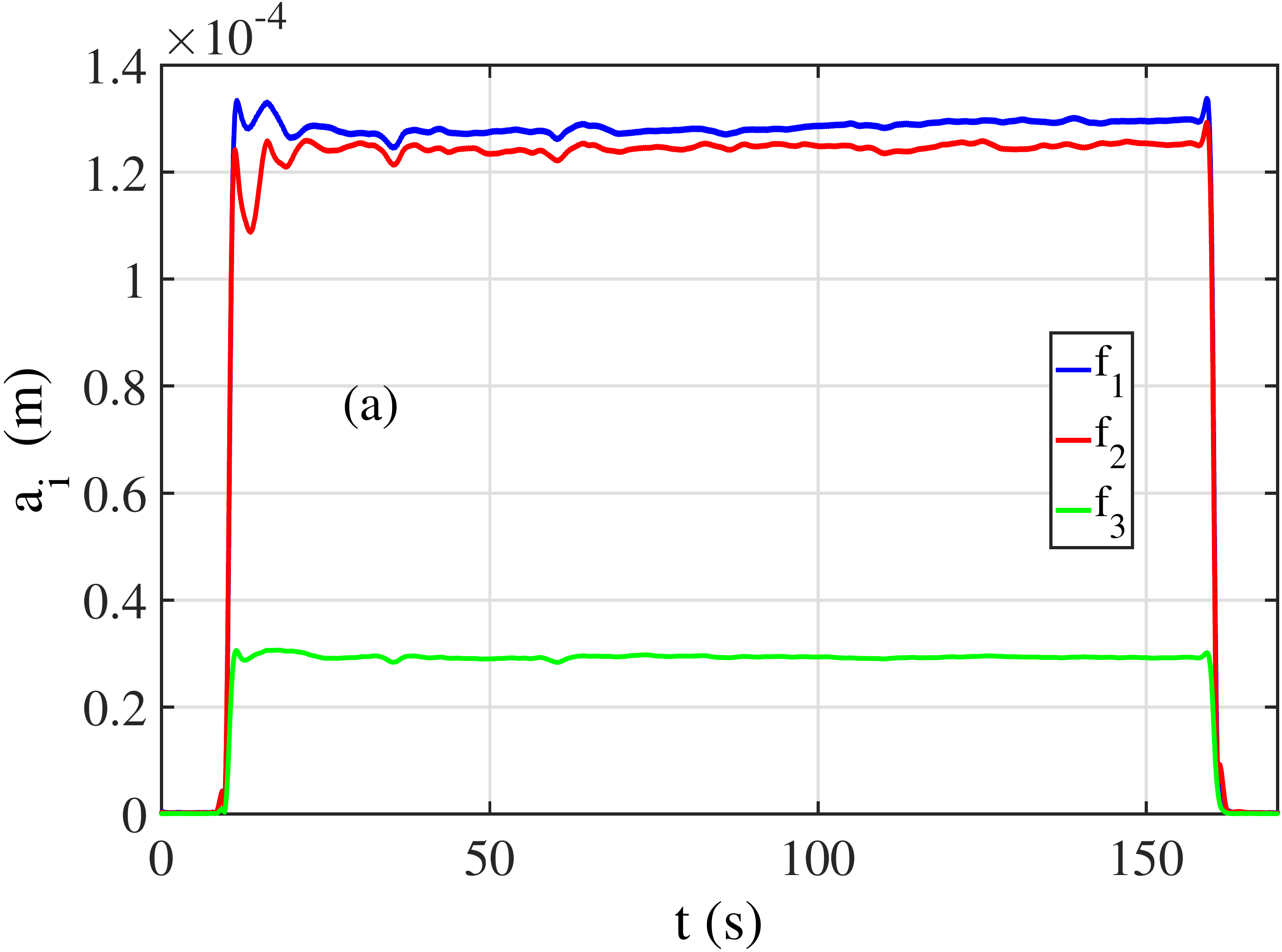} \hfill
 \includegraphics[width=0.48\columnwidth]{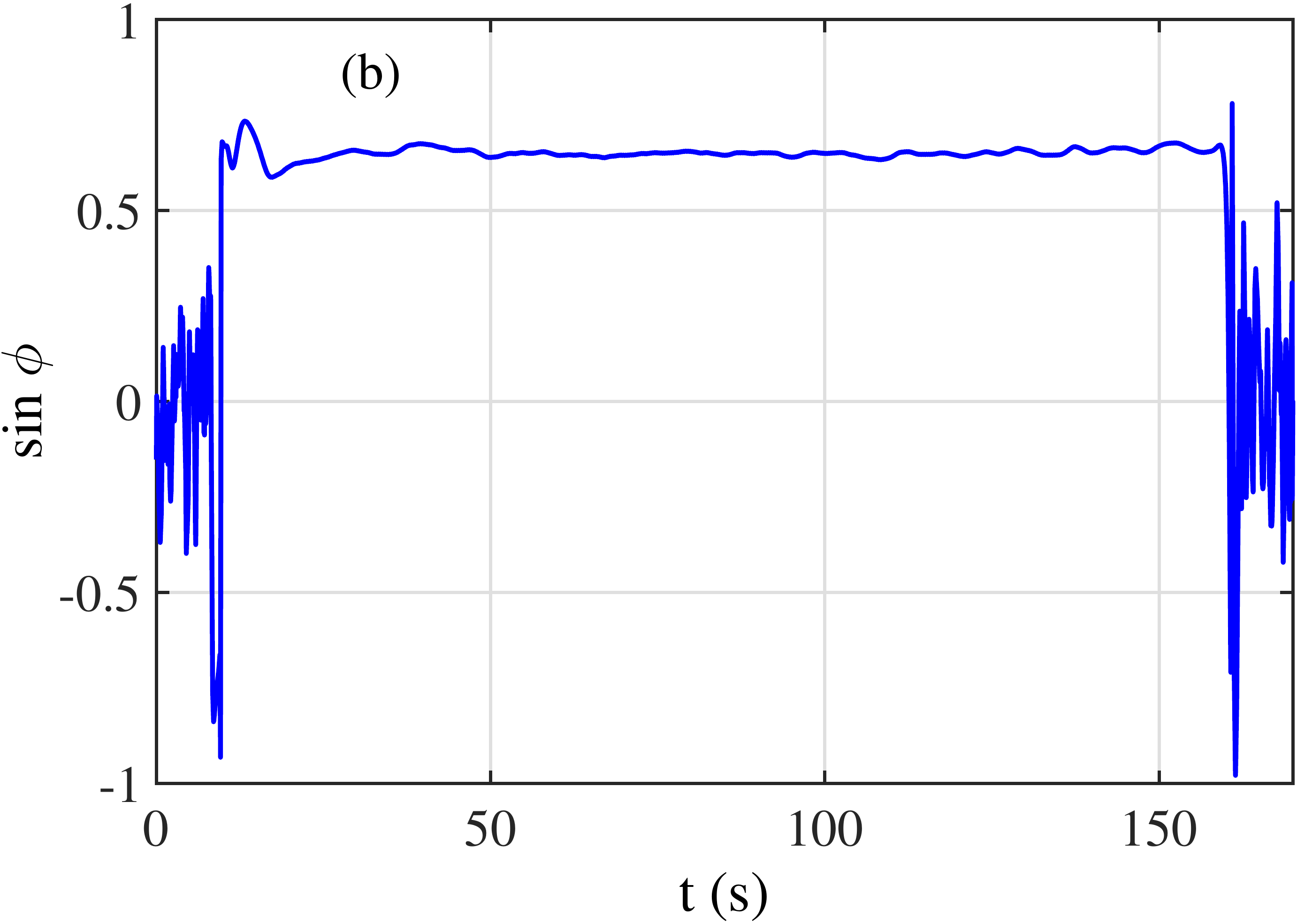} 
  \caption{(a) Amplitudes $a_i$ as a function of the time for $f_1=15\,$Hz, $f_2=18\,$Hz and $\alpha_{1\,2}=40\,^\circ$. (b)  $\sin \phi$ versus $t$ with $\phi=\phi_1+\phi_2-\phi_3$, $\phi_i$ is the phase of the wave $i$ measured using a Hilbert transform. A phase locking of the triad at {$\sin \phi \approx 0.65$} is observed. The measurement is performed with a laser vibrometer at a point located at $90\,$mm to both wavemakers, {which corresponds to a distance to the point $O$, $\xi \approx 70\,$mm (see point $M$ in Fig.~\ref{schema3ondes} (b)).}}
    \label{a1a2a340degampl9}
       \end{center}
 \end{figure}

\begin{figure}[!h]
 \begin{center}
  \includegraphics[width=0.48\columnwidth]{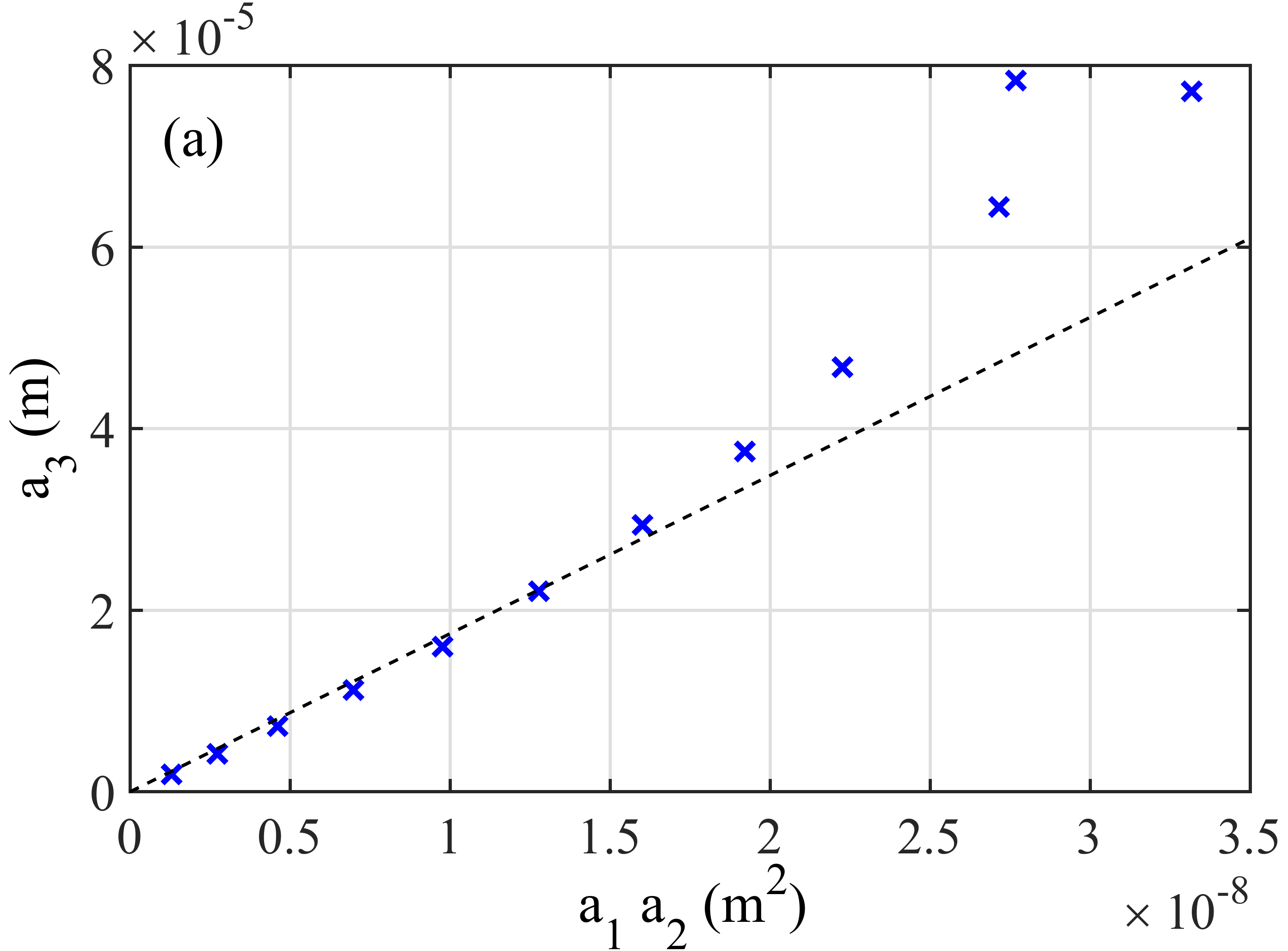} \hfill
 \includegraphics[width=0.48\columnwidth]{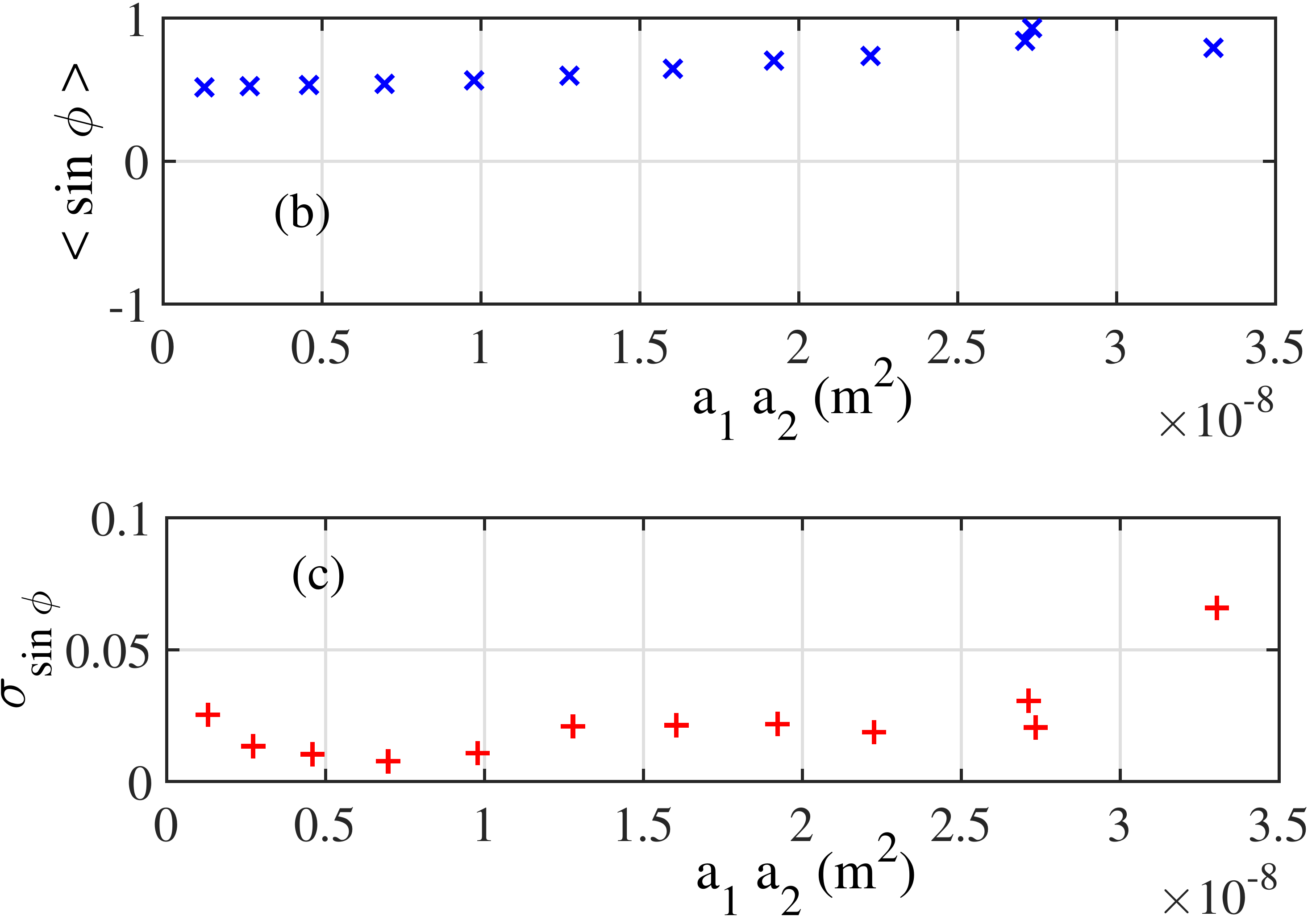} 
  \caption{(a) Amplitude of the daughter wave $a_3$ as a function of the product of the amplitudes of the mother waves $a_1\,a_2$, for $f_1=15\,$Hz, $f_2=18\,$Hz and $\alpha_{1\,2}=40\,^\circ$. A proportional relation is observed for sufficiently small mother wave amplitudes. (b) mean value of $\sin \phi$ versus $a_1\,a_2$, with $\phi=\phi_1+\phi_2-\phi_3$. (c) standard deviation of  $\sin \phi$ versus $a_1\,a_2$. For $a_1\,a_2 < 3\,10^{-8}$\,m$^2$ the level of fluctuations $\sigma_{\sin \phi}$ is small compared to the $|\langle\sin \phi \rangle|$, which evidences a robust phase locking of the triad at {$\sin \phi \approx 0.65$\,. The measurements are performed with a laser vibrometer at a point located at $\xi \approx 70\,$mm (see point $M$ in Fig.~\ref{schema3ondes} (b)).}}
    \label{a3vsa1a240}
       \end{center}
 \end{figure}    
Local measurements using a laser vibrometer provide other evidences arguing for a non-linear interaction mechanism. We assume that the wave field can be written as: $\eta(x,y,t)=\sum_i^3 \, a_i\,\cos(\omega t - \mathbf{k_i}\cdot (x \, \mathbf{e_x}+y \, \mathbf{e_y}))$. After band-pass filtering the local signal around $f_i$, the amplitudes of each components of the triad $a_i$ and the respective phase $\phi_i$ can be extracted using a Hilbert transform~\cite{Haudin2016}. {To avoid possible phase jumps, we compute the sine of the total phase $ \phi= \phi_1+\phi_2-\phi_3$. The evolutions of $a_i$ and $\sin \phi$} are plotted versus time in Fig.~\ref{a1a2a340degampl9} (a) and (b) again for $f_1=15\,$Hz, $f_2=18\,$Hz and $\alpha_{1\,2}=40\,^\circ$. When the wave $3$ is generated, a phase locking is evidenced around ${\sin \phi \approx 0.65}$\,. But the locking value departs from the value $\phi=\pi/2$ predicted when $\alpha_{1\,2}=\alpha_{1\,2\,r}$  and measured previously~\cite{Haudin2016}. 
 By measuring the temporal average of the $a_i$ for various forcing level, we find in Fig.~\ref{a3vsa1a240} (a), that the daughter wave amplitude $a_3$ is proportional to the product of the mother waves $a_1\,a_2$, for sufficiently small mother wave amplitudes (approximately for mother amplitude below $0.15$\,mm). {On this range,} the phase locking of the triad {$\sin \phi \approx  0.65$} remains robust with the change of $a_1\,a_2$ as seen in Fig.~\ref{a3vsa1a240} (b) by plotting the temporal average of {$\sin \phi$}. The fluctuations of {$\sin \phi$} estimated by its standard deviation in Fig.~\ref{a3vsa1a240} (c) are one order of magnitude smaller. {The scaling $a_3 \propto a_1\,a_2$ at low enough forcing amplitude remains valid for other measurement locations inside the region where the mother waves cross each other. However, the value of $a_3$ is spatially dependent like in the classic case~\cite{Haudin2016}. In contrast with this last case, the measured phase-locking value of $\phi$ depends on the sensor location.} The same qualitative observations have been found for other values of $\alpha_{1\,2}$ and also for another triad $f_1=16\,$Hz, $f_2=23\,$Hz and $f_3=f_1+f_2=39\,$Hz. The proportionality of the daughter wave amplitude with the product of mother wave amplitudes have been also tested for various choices of the mother wave frequencies using a capacitive local probe. This characteristic and the phase locking of the triad phase are strong arguments to attribute the daughter wave existence to a three-wave interaction mechanism even when $\alpha_{1\,2} \neq \alpha_{1\,2\,r}$.

\section{Model of forced three-wave interaction}
\label{Model3Wave}

\subsection{{Derivation}}
In Sect.~\ref{Firstevidence}, one observes a daughter wave created by the nonlinear interactions between mother waves obeying the resonant conditions and significantly outside the dispersion relation. We interpret this surprising observation as a forced three-wave interaction. The concept of forced waves interactions was mentioned in the pioneering work of F. P. Bretherton~\cite{Bretherton1964} investigating fundamentally resonant interactions in a model dealing with a one-dimensional conservative wave equation with a quadratic nonlinearity. These forced oscillations do not satisfy the linear dispersion relation and have a bounded magnitude, which is small compared to a resonant interaction verifying the dispersion relation. 
{Here,} we aim to determine thus the forced response of the liquid free surface in presence of two mother capillary-gravity waves interacting nonlinearly. To address this problem, we adapt the computations establishing the three-wave resonant interaction mechanism in weakly nonlinear regime for capillary-gravity waves. The classic methods can be a perturbation of the direct equations~\cite{McGoldrick1965,Case1977} or a development of a Lagrangian~\cite{Simmons1969} or of a Hamiltonian~\cite{Zakharov1967} function describing inviscid wave propagation. {Recently, resonant triad interactions have been revisited using a Hamiltonian formulation for water waves~\cite{Chabane2018}, demonstrating among others the possibility of resonant interactions without energy exchange, when the phase difference between the triad components has a specific value.} Here, following the approach of Case and Chiu~\cite{Case1977}, we consider the deep water limit and the hypothesis of a potential flow. Then, we only keep the first order in the nonlinear development, valid for weak nonlinearity. The details of the calculations are provided in appendices \ref{Modelderivation} and \ref{Amplitudeequation3}. The quadratic nonlinearities induce through the product of the two mother waves an excitation at the angular frequency $\omega_3=\omega_1+\omega_2$ and at the wavenumber $\mathbf{k_p}=\mathbf{k_1}+\mathbf{k_2}$. We thus express the forced dynamics of the daughter wave $3$ using the complex formalism $\underline{\eta_3}=B_3(\xi,t)\,\e^{i(\omega_3 t - k_p \xi)}+\mathrm{c.c.}$ along its propagation direction $O_\xi$ (see Fig.~\ref{schema3ondes}).  An amplitude equation for the complex envelope $B_3$ can be then found (Eq.~(\ref{B3ap}) in the Appendix) {in which the envelope propagates at the velocity $v_p$}: 

\begin{equation}
 \dfrac{\partial B_3}{\partial t}+v_p\,\dfrac{\partial B_3}{\partial \xi}-i \left(\dfrac{\omega_p^2-\omega_3^2}{2\,\omega_3}\right)B_3=\dfrac{-i}{2}\,\left( \dfrac{k_p}{\omega_3} T_A-T_B\right) \,B_1\,B_2
\end{equation}
with $v_p= \dfrac{1}{2}\left(\dfrac{\omega_3}{k_p}+\dfrac{2 \gamma\,k_p^2}{\rho\,\omega_3} \right)$ the {envelope} velocity of the daughter wave $3$,  $\omega_p=\sqrt{g\,k_p+\dfrac{\gamma}{\rho}\,k_p^3}$ and the angular frequency given by the linear dispersion relation for the wavenumber $k_p$. The left hand side gathers the temporal variation and the spatial variation of the envelope $B_3$ as well as an oscillation term due to the pure imaginary factor of $B_3$, whose period diverges in the case $\alpha_{1\,2}=\alpha_{1\,2\,r}$. The right hand side is the forcing term proportional to the mother wave amplitude product $B_1\,B_2$, the complex amplitudes of the mother waves. 
This equation has been obtained with the hypothesis of inviscid fluid. However, experimentally, capillary waves are known to dissipate on few centimeters by viscosity. Similarly to other experimental studies of three-wave interactions for gravity-capillary waves~\cite{McGoldrick1970,Henderson1987_1,Haudin2016}, a linear dissipation term is added characterized by the decay rate $\delta_3$. The precise value of $\delta_3$ can be complicated to know~{as a water surface is commonly contaminated by surfactants present in the atmosphere. An enhanced wave dissipation is indeed caused} by the variation of concentration of the surfactants trapped at the water/air interface~\cite{Alpers1989}. However, for capillary waves whose frequency is larger than the gravity-capillary crossover around $f=13\,$Hz for water, $\delta_3$ is well approximated by the {inextensible film} model~\cite{Henderson1987_1,HendersonSegur2013,Deike2012,Haudin2016}, which by considering a non-slipping {horizontal} velocity at the free-surface gives $\delta_3=\sqrt{2\,\nu\,\omega_3}\,k_3/4$~\cite{Lamb1932,VanDorn}, with $\nu$ the kinematic viscosity of the liquid. The evolution of the complex amplitude of the wave $3$ is then given by (see Eq.~(\ref{EqevolB3b})):

\begin{equation}
 \dfrac{\partial B_3}{\partial t}+v_p\,\dfrac{\partial B_3}{\partial \xi}+ \left[ \delta_3 - i \left(\dfrac{\omega_p^2-\omega_3^2}{2\,\omega_3}\right) \right] B_3=\dfrac{-i}{2}\,\left( \dfrac{k_p}{\omega_3} T_A-T_B\right) \,B_1\,B_2
 \label{EqevolB3}
\end{equation}

About the forcing term in the right-hand side, we note that the product of  $B_1$ and $B_2$ is non-null, only in the interaction area, where the mother wave beams are crossing. $T_A$ and $T_B$ are the interaction coefficients depending on the frequencies, wavenumbers and directions of the mother waves: $T_A =\dfrac{\omega_1 \,\omega_2}{k_1\,k_2 } (k_1\,k_2 - \mathbf{k_1} \cdot \mathbf{k_2}) + \omega_1^2+\omega_2^2$ and $T_B = \omega_1 \left(k_1+\dfrac{\mathbf{k_1} \cdot \mathbf{k_2}}{k_1} \right)+\omega_2 \left(k_2+\dfrac{\mathbf{k_1} \cdot \mathbf{k_2}}{k_2}\right) $. In the classic case, where $\alpha_{1\,2}=\alpha_{1\,2\,r}$, we have $k_p=k_3$ and $\omega_p=\omega_3$. Then it can be shown that Eq.~(\ref{EqevolB3}) becomes equivalent to the amplitude and phase evolution equations of the daughter wave $3$ for a resonant triad~\cite{Simmons1969,Case1977,Henderson1987_1}.\\

The resolution of Eq.~(\ref{EqevolB3}) in the stationary regime for constant mother wave amplitudes provides a physical insight of the physics of forced interactions. {We introduce $K=\dfrac{1}{2}\,\left( \dfrac{k_p}{\omega_3} T_A-T_B\right)$}. Then, with the boundary condition $B_3(\xi=0)=0$, one obtains:
 \begin{equation}
B_3(\xi)=\dfrac{K\,B_1\,B_2}{\dfrac{\omega_p^2-\omega_3^2}{2 \omega_3}-i \, \delta_3} \,\left[ 1-\e^{-\dfrac{\delta_3}{v_p} \,\xi}\,\e^{i\dfrac{\omega_p^2-\omega_3^2}{2\omega_3\,v_p}\xi}\right] 
\label{eqB3x}
\end{equation}
The wave elevation at the frequency $f_3$ defined by  $\eta_3(\xi,t)=\frac{1}{2}\,\mathfrak{Re} (\underline{\eta_3} (\xi,t))=\frac{1}{2}\,\mathfrak{Re} \left( B_3(\xi) \,  \e^{i(\omega_3 t - k_p \xi)}\right)$ is then deduced:
 
 \begin{equation}
 \eta_3(\xi,t) =\frac{1}{2}\, \mathfrak{Re} \left(\e^{i \omega_3 t}\dfrac{K\,B_1\,B_2}{\dfrac{\omega_p^2-\omega_3^2}{2 \omega_3}-i \, \delta_3} \,\left[\e^{-i\, k_p \xi}- \e^{-\dfrac{\delta_3}{v_p} \,\xi}\,\e^{i\,\left( \dfrac{\omega_p^2-\omega_3^2}{2\omega_3\,v_p}-k_p\right)  \xi} \right] \right)
 \label{eta3full}
\end{equation}
The free surface response time oscillating at $\omega_3$ contains a forced response at the wavenumber $k_p$ and {a viscous-damped transient response} $\mathrm{exp}\left(-\dfrac{\delta_3}{v_p} \,\xi\right)$, whose corresponding wavenumber is close to $k_3$ if {$k_p-k_3 \ll k_3$}. This transient can be then interpreted as the free response of the interface. In that limit, the evolution of the mode $3$ is identical to the one obtained for a quasi-resonant interaction, where $\mathbf{k_3}=\mathbf{k_1}+\mathbf{k_2}+\mathbf{\delta_k}$. Therefore, in the case of a forced interaction, the free surface excited at $k_p$ responds at  $k_p$ (outside the linear dispersion relation) and at the wavenumber $k_3$ (on the linear dispersion relation at $\omega_3$). Far from the origin at $\xi=0$, only remains the forced response whose amplitude is given by the factor $(K\,B_1\,B_2)/\left({\dfrac{\omega_p^2-\omega_3^2}{2 \omega_3}- \, i \,\delta_3}\right)$. The maximum of this factor is obtained for $\omega_p=\omega_3$, \textit{i.e.} $k_p=k_3$, which corresponds to a usual resonance phenomenon with the divergence of amplitude in absence of dissipation. For an angle between the mother waves close to $\alpha_{1\,2\,r}$, the behavior of the free surface is thus analog to a classic forced harmonic oscillator. {Finally, at the angular resonance $\alpha_{1\,2}=\alpha_{1\,2\,r}$ where $k_p=k_3$, $\omega_p=\omega_3$ and $v_p=v_{g\,3}$, we note that Eq.~(\ref{eta3full}) identifies to the solution found previously in Haudin et al.~\cite{Haudin2016} for real amplitudes, where the viscous damping saturates the nonlinear interaction:
 \begin{equation}
 \eta_{3\,r}(\xi,t) =\frac{1}{2}\, \mathfrak{Re} \left(\e^{i \omega_3 t-i\, k_3 \xi}\,\dfrac{i\,K\,B_1\,B_2}{ \delta_3} \,\left[1- \e^{-\dfrac{\delta_3}{v_{g\,3} \,\xi}} \right] \right) \quad \quad \mathrm{when} \quad \alpha_{1\,2}=\alpha_{1\,2\,r}
 \label{eta3full_r}
\end{equation}}

\begin{figure}[h!]
\centering
\includegraphics[width=0.46\columnwidth]{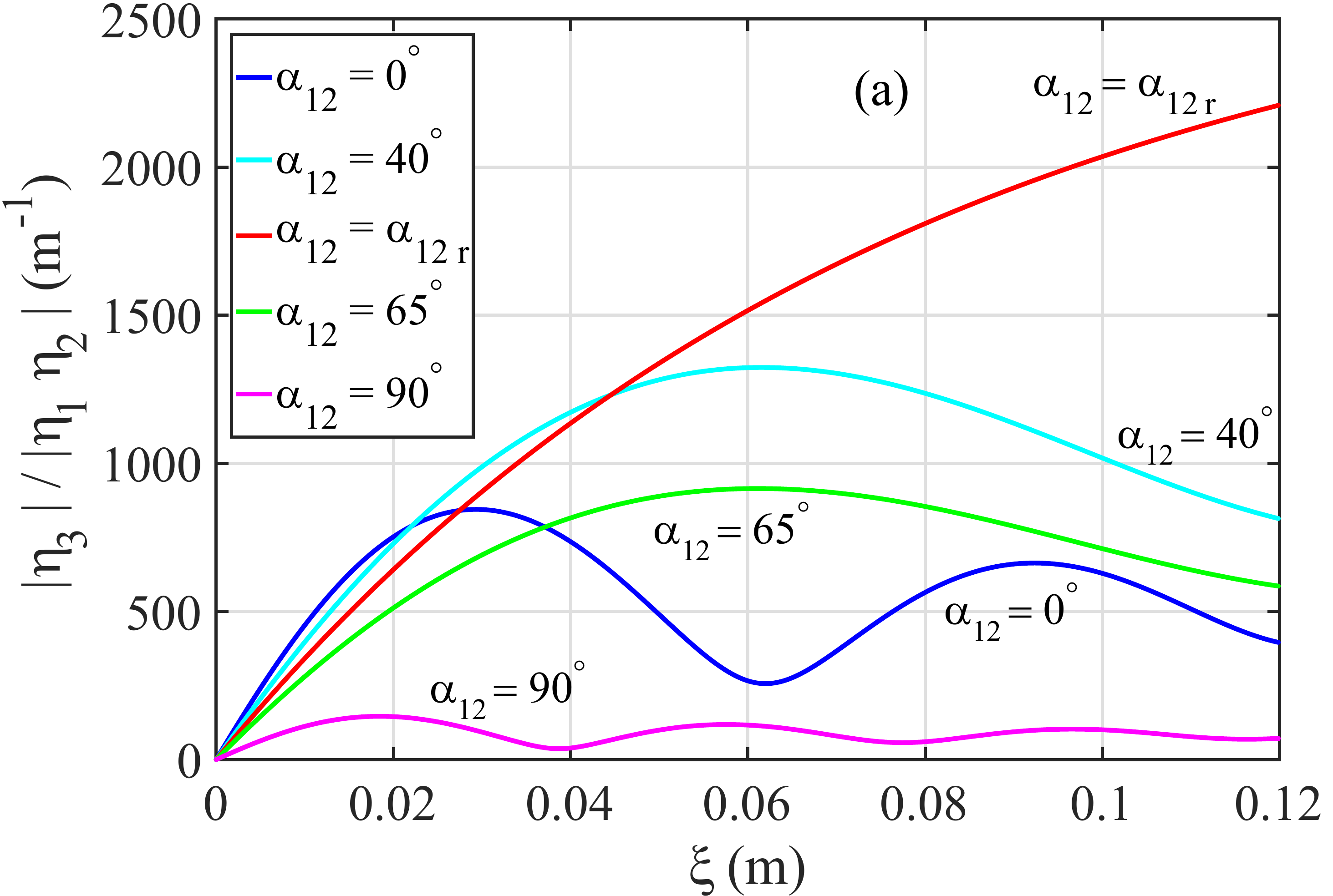}\hfill
\includegraphics[width=0.46\columnwidth]{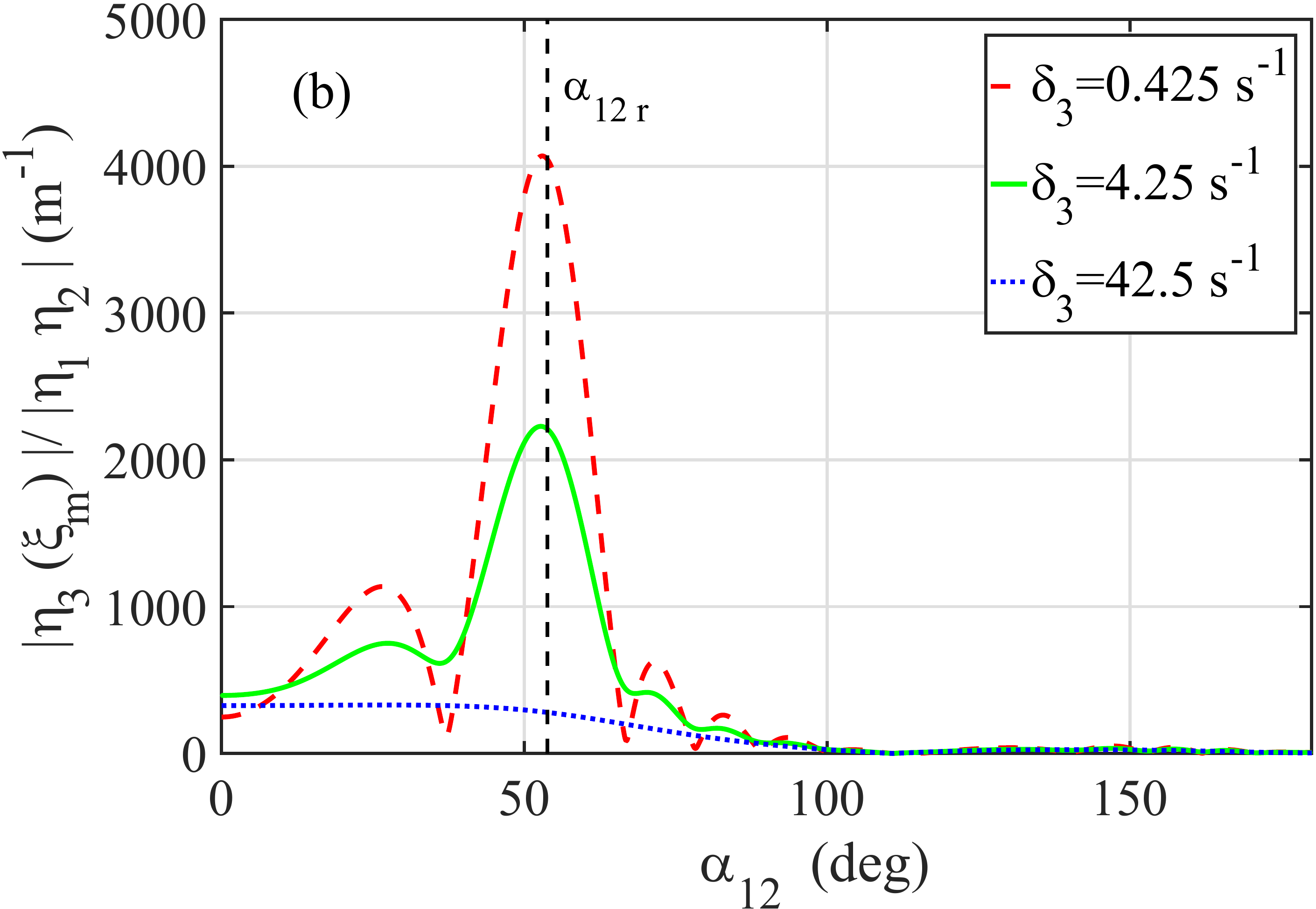}
\caption{Prediction of the model of forced three-wave interaction. (a) normalized amplitude $|{\eta_3} (\xi) / |\eta_1 \, \eta_2 |$ as a function of $\xi$, for various values of the angle $\alpha_{1\,2}$ between the two mother waves from Eq.~(\ref{eta3full}). (b) normalized amplitude $|\underline{\eta_3}(\xi_m)|/ |\eta_1 \, \eta_2 |$ evaluated in $\xi_m=0.12$\,m as a function of $\alpha_{1\,2}$ for various values of the dissipation rate ({$\delta_3=4.25$\,s$^{-1}$} is the expected dissipation at $f_3=33\,$Hz in the experiments). The curve for negative values of $\alpha_{1\,2}$ is symmetric to the displayed one, due to the parity of the interaction coefficients with $\alpha_{1\,2}$.}
\label{eta3reso}
\end{figure}

\subsection{{Prediction of the daughter wave amplitude with respect to the relevant parameters}}
{In the general case $\alpha_{1\,2} \neq \alpha_{1\,2\,r}$}, the analytic solution can be easily plotted  with the parameters corresponding to the triad $f_1=15$\,Hz, $f_2=18$\,Hz and $f_3=33\,$Hz, {which are given in Table~\ref{Tab1}. We have checked that variations of the surface tension $\gamma$ and of the dissipation coefficient $\delta_3$ in the range typically occurring in the experiments ($\gamma \in [50,74]$\,mN/m and $\delta_3 \in [4,5]\,$s$^{-1}$) do not change qualitatively the results presented in this part}.

The evolution of the normalized daughter wave amplitude with the distance of propagation $\xi$ is plotted in Fig.~\ref{eta3reso} for few values of the angle $\alpha_{1\,2}$. When $\alpha_{1\,2}=\alpha_{1\,2\,r}$, the daughter wave grows to reach a saturation level set by the decay rate $\delta_3$ as described in our previous work~\citep{Haudin2016}. For the other values, the wave $3$ displays envelope modulations similar to acoustic beats, with a typical wavenumber of order $|k_p-k_3|$. At short distance, we note that the daughter wave grows faster for small $\alpha_{1\,2}$, because  $\alpha_{1\,2}=0$ is a local maximum of the interaction coefficient $|K|$. At longer distance, the {daughter wave obtained for $\alpha_{1\,2\,r}$} dominates. An angular resonance can be indeed evidenced by plotting the rescaled daughter wave amplitude as a function of $\alpha_{1\,2}$ in Fig.~\ref{eta3reso} (b) for a particular distance like for example $\xi_m=0.12$\,m, the half of the experimental tank diameter. The response is indeed maximal at the vicinity of $\alpha_{1\,2\,r}$, however for this level of dissipation, the response is not negligible outside the angular resonance, especially for $\alpha_{1\,2} < \alpha_{1\,2\,r}$. When $\delta_3$ is taken smaller, the bandwidth of the resonance decreases and the peak becomes higher as displayed in  Fig~\ref{eta3reso} (b). On the contrary, for high values of the dissipation, the resonant behavior disappears. {This angular resonance can be physically interpreted by analogy with a harmonic oscillator. Here, the free-surface oscillating at $\omega_3$ is excited at $k_p$, which differs from the wavenumber obeying to the dispersion relation $k_3$. This model shows that the daughter wave has a maximal amplitude when $k_p \approx k_3$, \textit{i.e.} $\alpha_{1\,2}= \alpha_{1\,2\,r}$.
As the linear dispersion relation corresponds to the response of the free-surface in absence of spatially extended forcing, the angular resonance occurs when the forcing wavenumber $k_p$ is close to the wavenumber $k_3$ of the free regime.}

\begin{figure}[h!]
\centering
\includegraphics[width=0.32\textwidth]{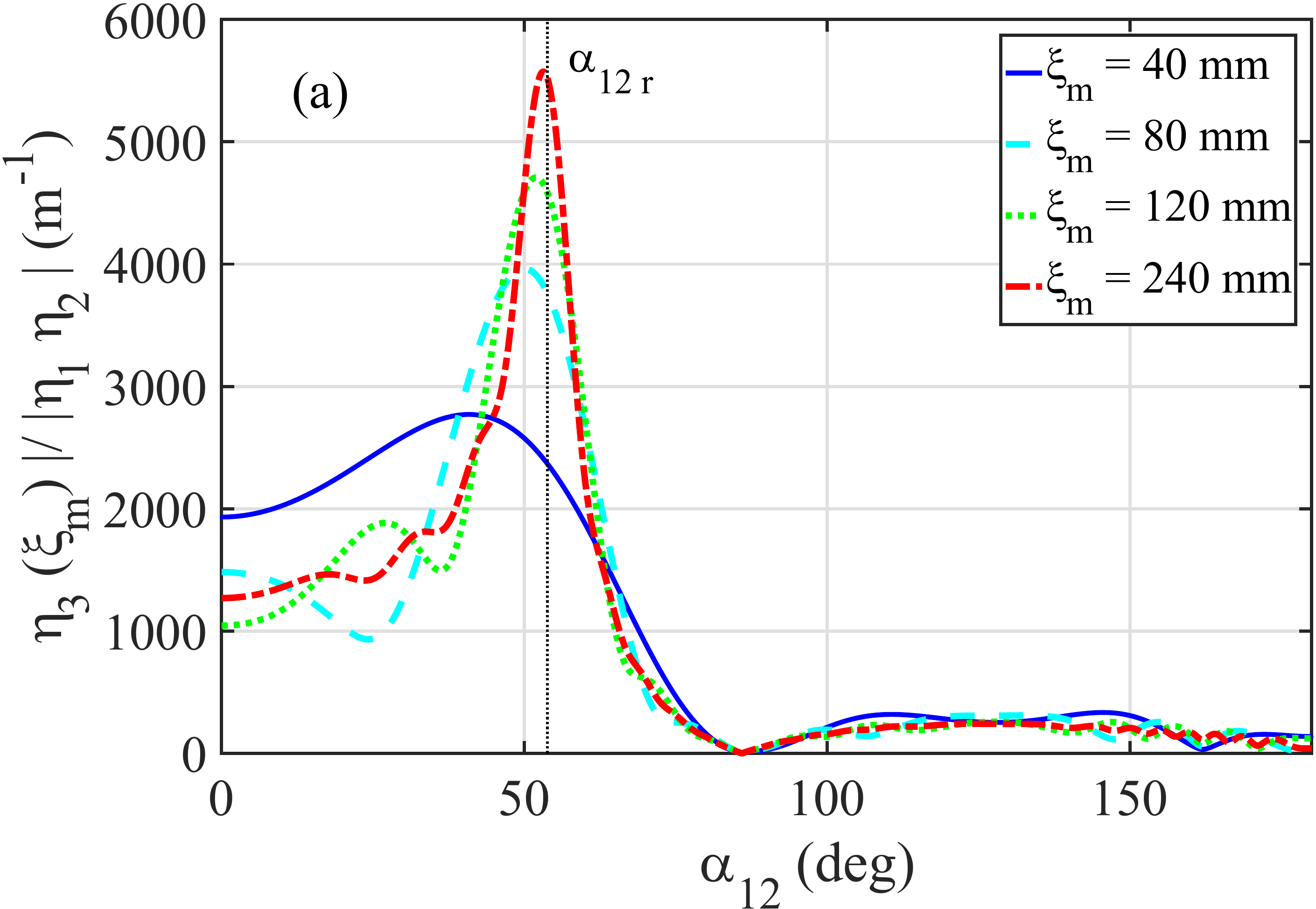}\hfill
\includegraphics[width=0.32\textwidth]{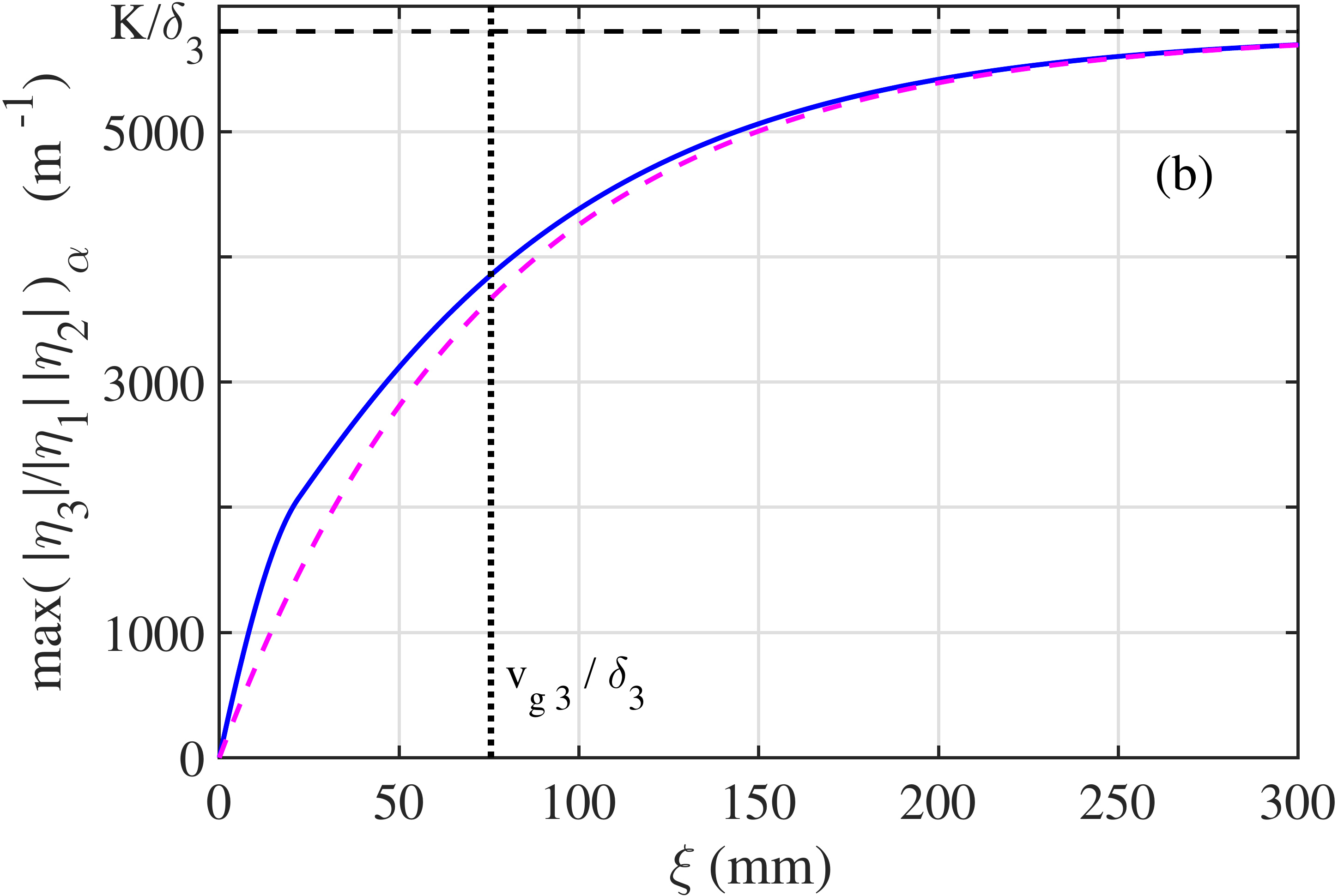} 
\includegraphics[width=0.32\textwidth]{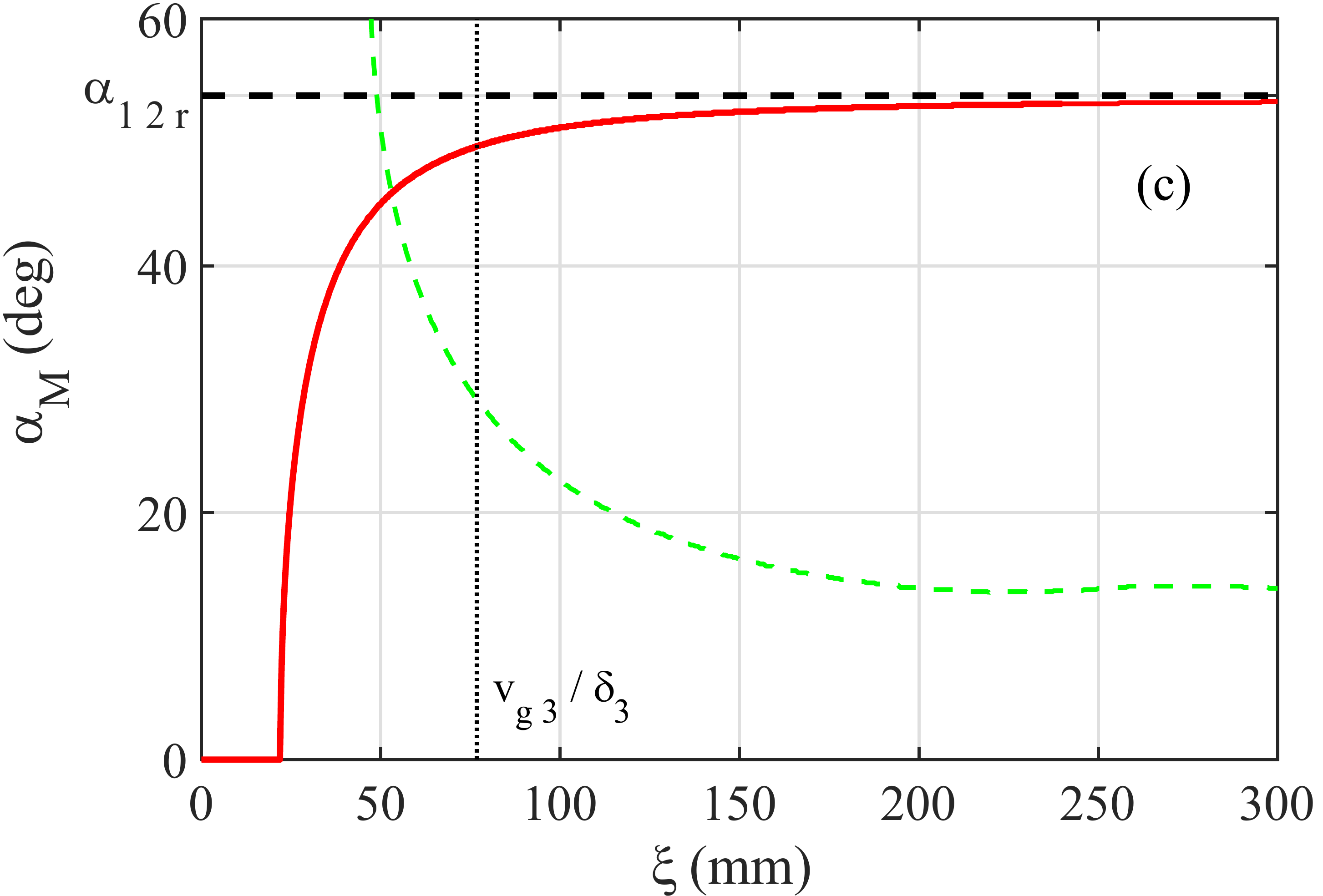}
\caption{(a) amplitude of the daughter wave $|\underline{\eta_3}(\xi_m)|$  (normalized by $|\eta_1\,\eta_2|$) as a function of $\alpha_{1\,2}$ for various values of the propagation distance $\xi_m$ for a dissipation rate  $\delta_3=4.25$\,s$^{-1}$.  (b) {for each value of the propagation distance $\xi$, the maximum of the curve  $|\underline{\eta_3}(\xi)|/(|\eta_1\,\eta_2|)$ versus $\alpha_{12}$ (see (a)) is reported as a function of $\xi$ (blue line).} Magenta dashed line, $K/\delta_3\,(1-\mathrm{exp}(-\xi\,\delta_3/v_{g\,3}))$ versus $\xi$, {solution when $\alpha_{1\,2}=\alpha_{1\,2\,r}$,  Eq.~(\ref{eta3full_r})}.  (c) red line, $\alpha_M$ position in $\alpha_{1\,2}$ of the previous maximum {(see Fig. 7 (a))} as a function of $\xi$. For large enough values of  $\xi$, $\alpha_M \rightarrow \alpha_{1\,2\,r}$, when $\xi$ increases. The half bandwidth of this resonance $\Delta_\alpha$ (green dashed line) decreases to a finite value due to the dissipation $\delta_3$. }
\label{distreso}
\end{figure}

As the amplitude of the daughter wave evolves in space, the position of the observation point matters. In Fig.~\ref{distreso}, the rescaled amplitude of the daughter wave $|\underline{\eta_3}|/|\eta_1\,\eta_2|$ is plotted as a function of the angle between the two mother waves for the dissipation rate expected at the frequency $f_3=33\,$Hz. The response of the free surface is not very selective on the angle at short distance for $\alpha_{1\,2} < 90^\circ$. After some propagation distance, the resonance peak appears and becomes well defined around $\alpha_{1\,2\,r}$. For large angles  $\alpha_{1\,2} > 90^\circ$ the response is always weak. 
To characterize the linear response of the free surface as a function  of the angle $\alpha_{1\,2}$ and of the propagation distance $\xi$, the maximal value of the rescaled daughter wave amplitude  $|\underline{\eta_3}|$ is plotted as a function of the distance in Fig.~\ref{distreso} (b) and the corresponding angular position $\alpha_M$ is given in Fig.~\ref{distreso} (c). The bandwidth at half the maximum $\Delta_\alpha$ is also indicated in the same plot with a dashed line ($\Delta_\alpha$  is not well defined for too short distance). {According to Eq.~(\ref{eta3full}), the forced regime at $k=k_p$ is reached once the transient response has sufficiently decayed, \textit{i.e.} the daughter wave has propagated on a distance typically larger than the decay length $v_p/\delta_3 \approx v_{g\,3}/\delta_3 \approx 76\,$mm. In Fig.~\ref{distreso} (b) we observe that the maximum grows thus on a spatial scale corresponding to this distance. For $\xi >  v_{g\,3}/\delta_3$, we note that the evolution of the maximum with $\xi$ is well approximated by the solution when $\alpha_{1\,2}=\alpha_{1\,2\,r}$ given by Eq.~(\ref{eta3full_r}). It saturates then at a value given by the balance between the nonlinear forcing and the viscous dissipation $|K(\alpha_{1\,2\,r})|/\delta_3 $. In Fig.~\ref{distreso} (b),  we observe that the response of the free surface becomes more and more peaked around the resonant angle: as the propagation distance $\xi$ increases, the position of the maximum goes to $\alpha_{1\,2\,r}$ and the bandwidth $\Delta_\alpha$ decreases towards a finite value due to the viscous dissipation modeled by  the decay rate $\delta_3$. Therefore, the resonant response of the free surface requires a certain distance of order  $v_{g\,3}/\delta_3$ to be observed. In contrast at shorter distance, a significant response for angles $\alpha_{1\,2} \neq \alpha_{1\,2\,r}$ is possible.} Moreover as it can be seen in Fig.~\ref{distreso} (a), we note that even at long distance,  for a realistic level of dissipation for frequency slightly larger than the gravity-capillary cross-over, the response for $\alpha_{1\,2} < \alpha_{1\,2\,r}$ is non negligible compared to the peak of maximal response (typically a factor $1/5$ for $\alpha_{1\,2}  \approx 0^\circ$). \\

\begin{figure}[h!]
\centering
\includegraphics[width=0.45\textwidth]{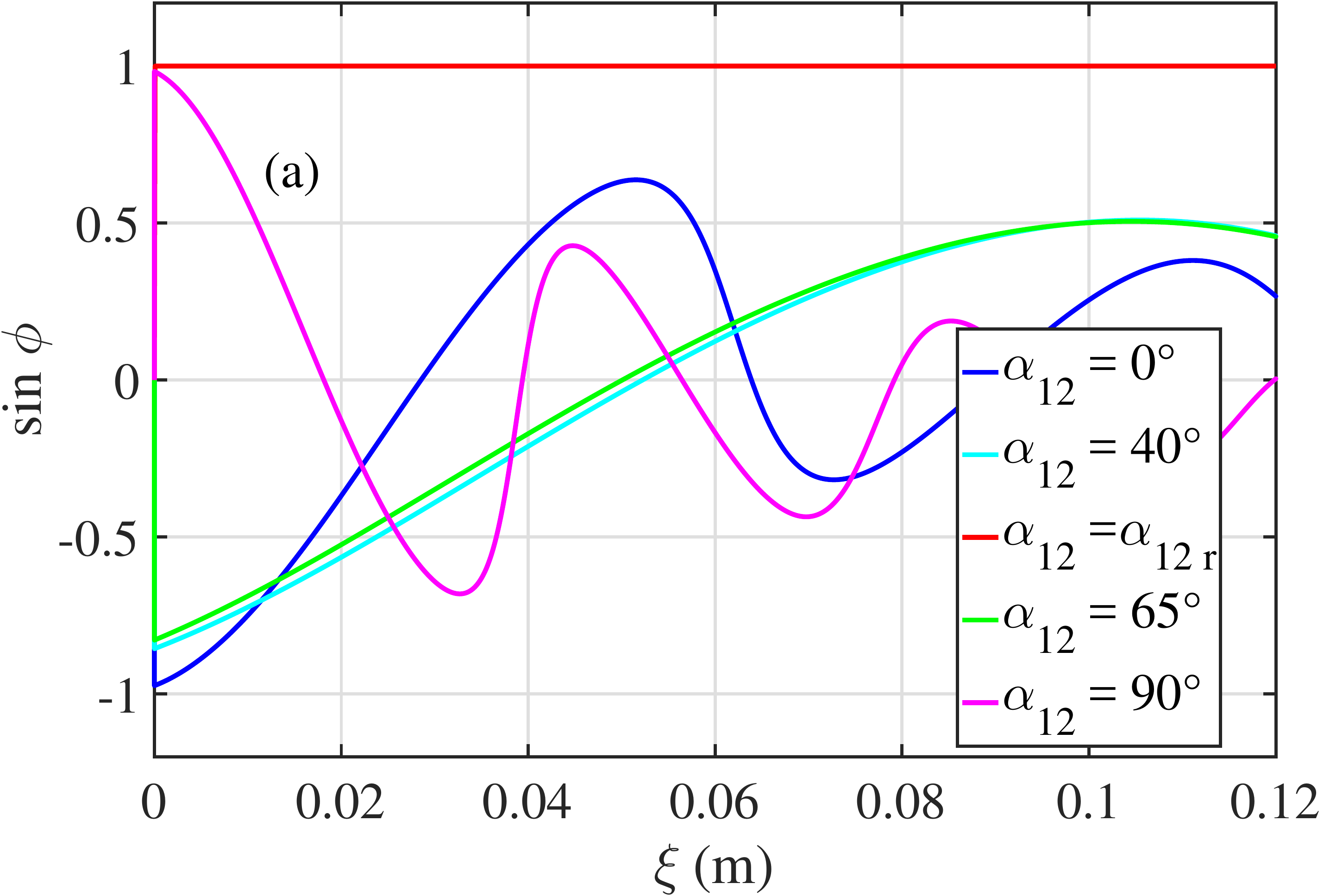}\hfill
\includegraphics[width=0.45\textwidth]{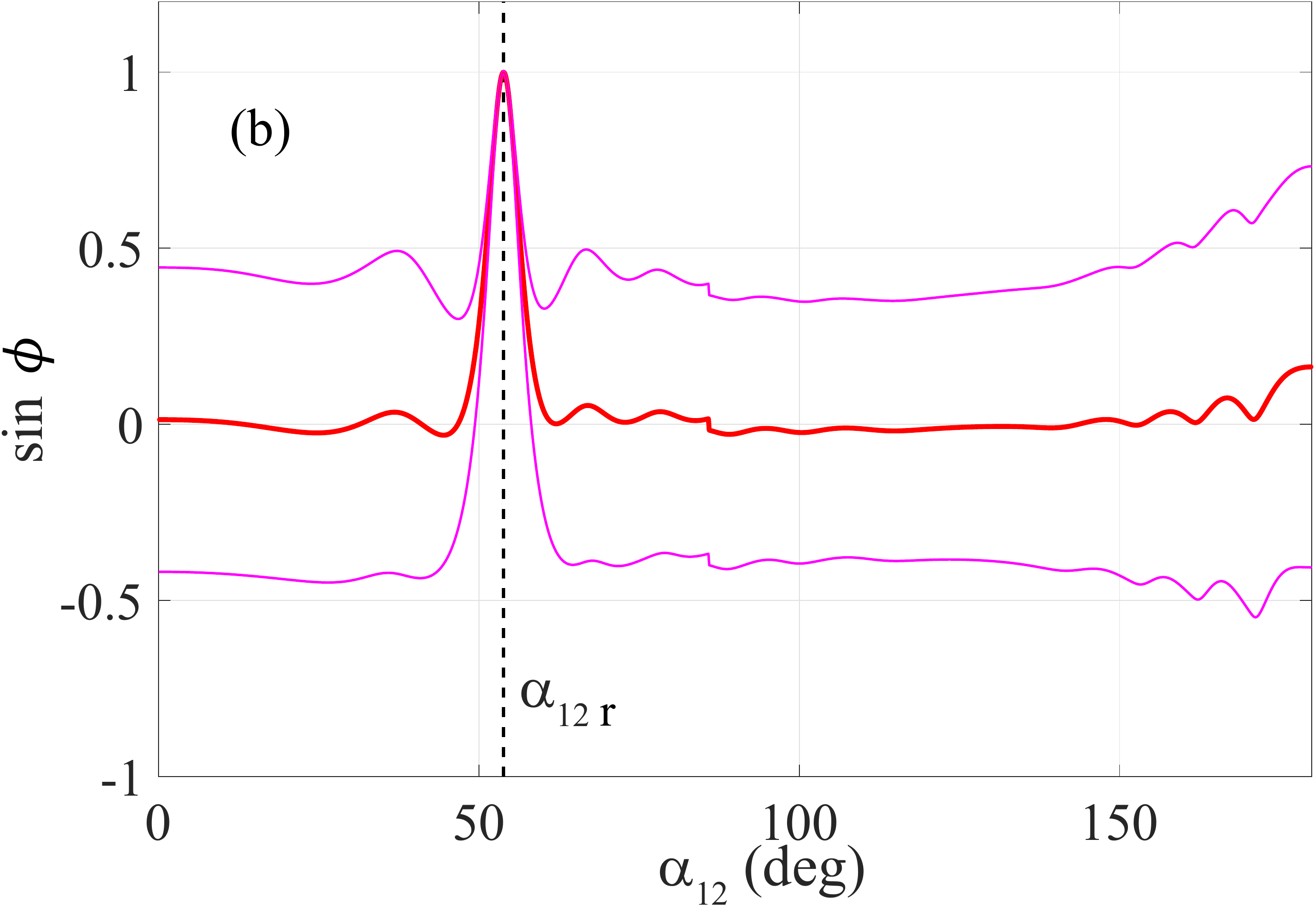}
\caption{{(a) sinus of the total phase $\phi=\phi_1+\phi_2-\phi_3$ as a function of $\xi$ in stationary regime for few values of the angle $\alpha_{1\,2}$ between the mother waves. When $\alpha_{1\,2}=\alpha_{1\,2\,r}$, the phase $\phi$ is locked to $\pi/2$, otherwise $\phi$ is spatially modulated, with a modulation wavelength increasing with the distance $\alpha_{1\,2}- \alpha_{1\,2\,r}$, to the classic case. (b) Red thick curve spatial average $\langle \sin \phi \rangle$ on the interval $\xi \in [0, 0.12]$\,m as a function of the angle $\alpha_{1\,2}$. To better visualize the fluctuations from the mean values,  $\langle \sin \phi \rangle \pm \sigma_{\sin \phi}$ are plotted in magenta thin curves, with  $\sigma_{\sin \phi}$ the standard deviation of $\sin \phi $ for $\xi \in [0, 0.12]$\,m. When $\alpha_{1\,2}\approx \alpha_{1\,2\,r}$, a clear phase locking is predicted at $\phi=\pi/2$ with vanishing fluctuations. For $\alpha_{1\,2}\neq \alpha_{1\,2\,r}$, $\langle \sin \phi \rangle \approx 0$ and the fluctuations are strong, showing that the total phase $\phi$ is spatially modulated. Similar results are expected for different but sufficiently long intervals. The dissipation of the daughter wave is set to $\delta_3=4.25\,$s$^{-1}$.}}
\label{phasetheo}
\end{figure}

\subsection{{Capture of the phase locking with the model}}
Using the complex spatial solution of the daughter wave $\underline{\eta_3}(\xi,t)$ in the stationary regime Eq.~(\ref{eta3full}), the total phase $\phi=\phi_1+\phi_2-\phi_3$ can be also computed as $\phi=    \mathrm{arg} \,(B_1\,\e^{i (\omega_1 t - \mathbf{k_1}\,\mathbf{\xi})})+\mathrm{arg}\, (B_2\,\e^{i (\omega_2 t - \mathbf{k_2}\,\mathbf{\xi})}\, ) -\mathrm{arg} \,(B_3 (\xi) \, \e^{i (\omega_3 t - k_p\,\xi)})$. As $\omega_1+\omega_2=\omega_3$ and $\mathbf{k_1}+\mathbf{k_2}=\mathbf{k_p}$, this last equation resumes in fact to $\phi=\mathrm{arg} \,(B_1)+\mathrm{arg} \,(B_2)-\mathrm{arg} \,(B_3 (\xi))$. {To avoid possible phase jumps, we discuss the results of this model by plotting the sine of the phase $\phi$, $\sin \phi$ in Fig.~\ref{phasetheo} (a). When $\alpha_{1\,2}=\alpha_{1\,2\,r}$, we predict as reported previously~\cite{Haudin2016} a phase locking of $\phi$ to the value $\pi/2$, \textit{i.e.} the phase of the daughter wave is set by the phases of the mother waves. This value $\pi/2$ maximizes the energy transfer from the mother waves to the daughter wave. When $\alpha_{1\,2}\neq \alpha_{1\,2\,r}$, the total phase $\phi$ is locked at a given position as observed experimentally in Fig.~\ref{a3vsa1a240} (b), but the locking value varies with $\xi$\,. $\sin \phi$ evolves indeed with $\xi$ as a non sinusoidal and damped oscillation whose the typical wavelength is larger than $2\pi/k_p$ and increases with the difference $\alpha_{1\,2}- \alpha_{1\,2\,r}$. An analog prediction has been reported for quasi-resonant four wave interactions for gravity surface waves~\cite{Bonnefoy2016}, for which the value of phase locking depends on the spatial distance to the wavemakers. To quantify the modulation of the phase, the spatial average of $\sin \phi$ along the first $120$\,mm is plotted in Fig.~\ref{phasetheo} (b) as a function of $\alpha_{1\,2}$. Outside the domain $\alpha_{1\,2} \pm 5^\circ$, the mean phase is essentially zero accompanied with strong fluctuations. Finally, in the example $\alpha_{1\,2}=40^\circ$, we note that the model predicts a phase locking value about $\sin \phi \approx 0.3$ for $\xi \approx 70\,$mm, whereas we report experimentally for this distance and this angle between the mother waves, a value of $\sin \phi$ between $0.5$ and $0.7$ in Fig.~\ref{a3vsa1a240} (b) for small enough mother wave amplitudes. This simple model predicts thus a phase relation of the daughter wave with the mother waves, but appears thus too simple to give accurately the local value of phase locking.}\\

\subsection{{Interpretation of forced interactions}}
{To conclude,} according to this simple model, this case of forced three-wave interaction verifying the resonant conditions but producing a daughter wave outside the linear dispersion relation is strongly analogous to the case of a non resonant interaction, where the daughter wave follows the dispersion but one of the resonant condition is not exact: $\delta_{\omega,NR} =   \omega_1 + \omega_2- \omega_3\neq 0$ or $\delta_{k,NR} = k_1 + k_2 - k_3  \neq 0$. Here, we have indeed $\delta_k=k_p - k_3= k_1 + k_2 - k_3$. Moreover, according to this model, outside the interaction zone, where at least one of the mother wave amplitude vanishes, the forcing at $k=k_p$ disappears and thus by continuity of the wave elevation, the daughter wave propagates at the frequency $f_3$ and at the wavenumber $k_3$ as this wavenumber corresponds to the free response of the free surface. Qualitatively, the forced three-wave resonant interaction mechanism is strongly analogous to the resonance of a classic damped oscillator. The forcing can excite the system at a frequency which differs from the eigen frequency of the oscillator (defined in absence of forcing), but the response is maximal at the vicinity of this eigen frequency and the bandwidth of this resonance increases with the dissipation level. An example of this situation is given by the sloshing motion of a fluid inside a tank, which is mechanically excited with a sinusoidal motion~\cite{Ibrahimbook}. The forcing frequency can also differ from the one determined by the motion of the largest eigen-mode, but a resonance occurs when these two frequencies are close.

Finally we note, that in order to keep an analytically easily solvable model, we have not taken into account the spatial variations of the mother waves which are due to the viscous dissipation and to the nonlinear energy pumping which generates the daughter wave. This assumption limits the application of our model to small daughter wave amplitudes and to systems whose size is smaller or comparable to the typical attenuation length of mother waves. Numerical simulations could be thus useful to determine quantitatively the influence of the spatial variations of mother waves.

\section{Experiments with different mother wave  angles $\alpha_{1\,2}$\,}

\subsection{Spatial analysis of the wave field}
In order to test the model presented in Sect.~\ref{Model3Wave}, another set of experiments has been performed with different values of  $\alpha_{1\,2}$ using the experimental device described in Sect.~\ref{ExpMethods} with the DLP method. The orientations of the wavemaker paddles are varied between the experiments. Due to the {volume occupied by the wavemakers during their motion}, the values of $\alpha_{1\,2} \lesssim 20^\circ$ are not accessible with our experimental setup. The angle between the mother waves $\alpha_{1\,2}$ is deduced from the measurement of the angle between the wavemaker paddles, which are imaged by the camera. The relative position of the interaction zone differs thus between the measurements. The amplitudes of the two mother waves are kept at a constant level for all measurements, sufficient to observe the generation of a daughter wave {with the DLP method for all values of $\alpha_{1\,2}$. These amplitudes correspond} to a steepness of order $0.07$ in the center of the tank{, which is however} quite high considering the hypothesis of weak nonlinearity in the theoretical model. Most of the measurements are performed for the triad $f_1=15$\,Hz, $f_2=18$\,Hz and $f_3=33$\,Hz. The use of a solution of intralipids for the liquid induces a smaller value of surface tension $\gamma = 55$\,mN.m$^{-1}$ compared to pure water. This value is found experimentally from the spatio-temporal spectrum $S_\eta(\omega,k)$ for this set of data.

First, we present in details the example with an angle $\alpha_{1\,2}=40.2 \approx 40 \,^\circ$, the results being similar for the other angle values. In order to identify the components of the wave field, {we can use a decomposition in the spatial Fourier space like in section~\ref{Firstexp}}. However, the dissipation of capillary waves is not negligible and the spatial decay of the waves must be taken into account. The wave-field is thus not homogeneous, which limits the relevance of an analysis relying on a spatial Fourier transform. To address this issue, we carry out an analysis in the spatial real-space enabling a quantitative comparison with our model.

\begin{figure}[h!]
\centering
\includegraphics[width=.32\textwidth]{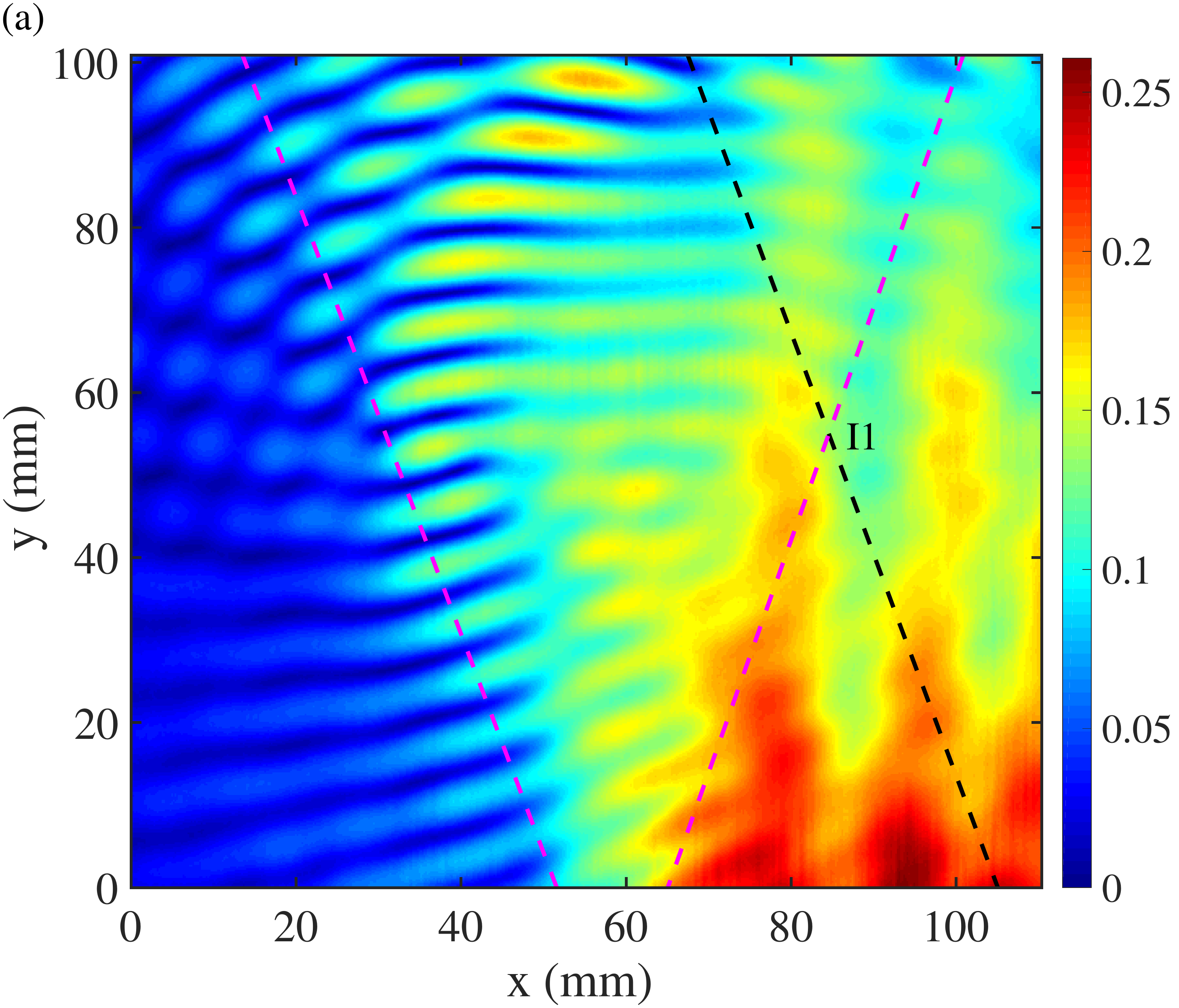} \hfill
\includegraphics[width=.32\textwidth]{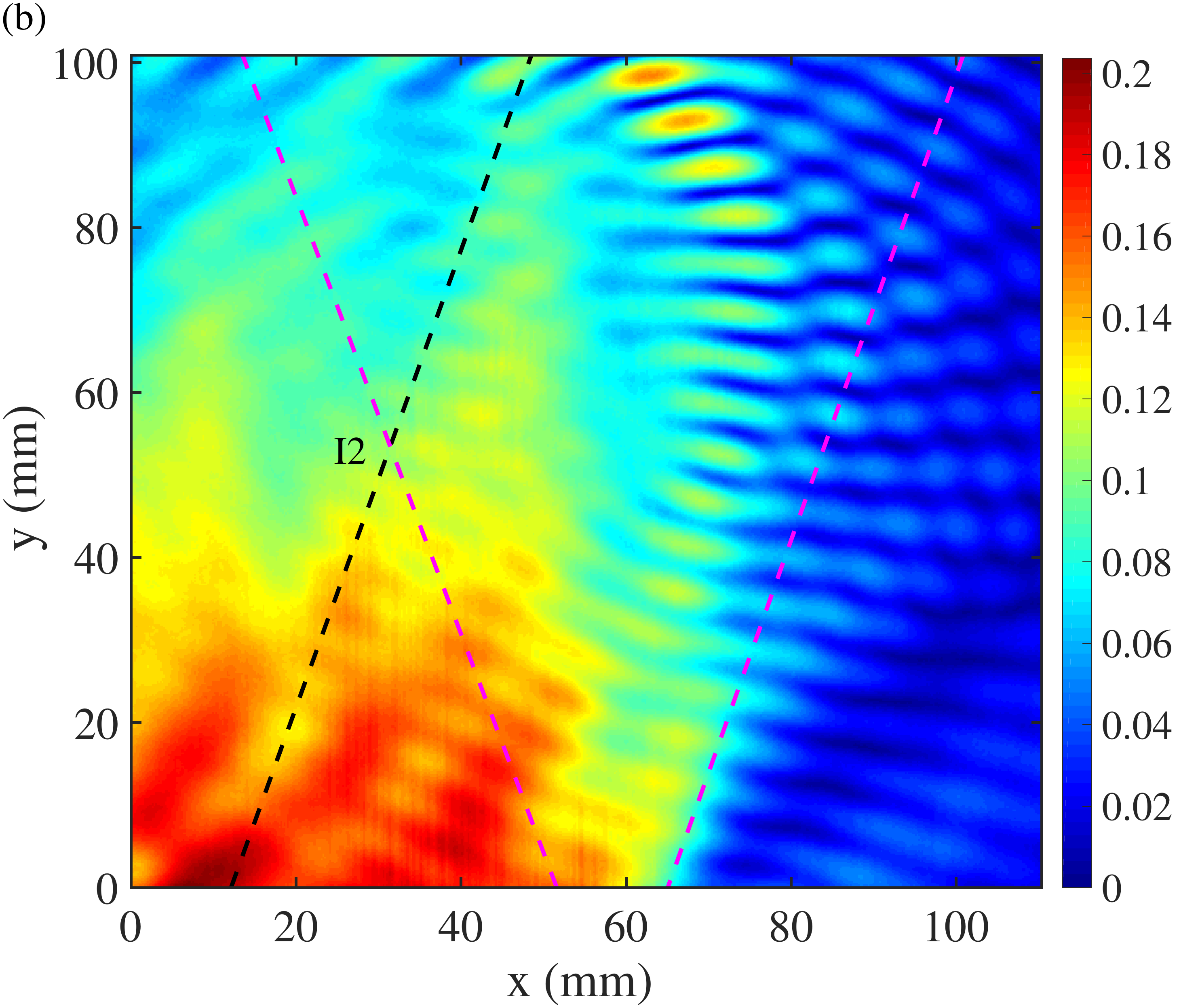}\hfill
\includegraphics[width=.32\textwidth]{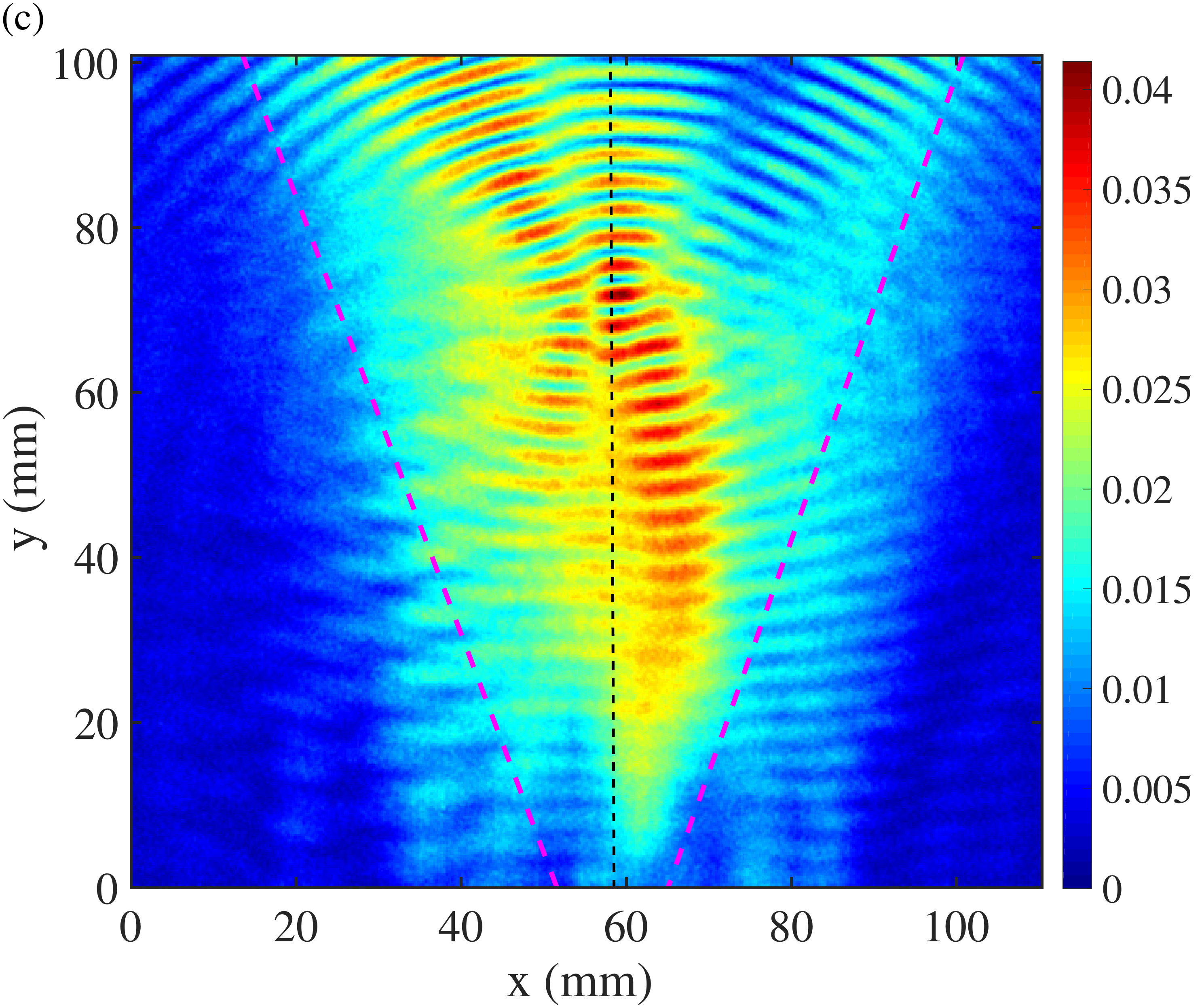} \hfill
\caption{(a), (b), (c) Spatial wave modes $a_i (x,y,f)$ for $f_1$ (a), $f_2$ (b) and $f_3$ (c) defined by computing a temporal Fourier transform $\tilde{\eta} (x,y,f) $ and taking the modulus $|\tilde{\eta}|$. Colorscale is in millimeter. {The interaction zone is depicted by magenta lines (see Fig.~\ref{schema3ondes}).  In (a) and (b), the black dashed lines are the lines starting from the middle of the corresponding wavemaker directed respectively by $\mathbf{k_1}$ and $\mathbf{k_2}$. $\mathrm{I_i} $ are the respective positions of entrance of the mother wave in the interaction zone. In (c), the black dashed line starting from the beginning of the interaction zone (point $O$) and directed by $\mathbf{k_p}=\mathbf{k_1}+\mathbf{k_2}$, corresponds to the axis $O_\xi$ (see Fig.~\ref{schema3ondes}). The daughter wave is essentially located inside the interaction zone between the two magenta lines.}}
\label{modespa}
\end{figure}

We perform a temporal Fourier transform to the surface deformation $\eta(x,y,t)$ (in stationary regime) and take the square of the modulus to get the spectrum $S_\eta(x,y,f)$. The amplitude of the wave modes $a_i$ at the frequency $f_i$ is defined by $a_i (x,y) =\sqrt{2}\,\left( \int_{f_i-\delta_f}^{f_i+\delta_f}\, S_\eta(x,y,f) \,\mathrm{d} f  \right)^{1/2}$, with $\delta_f=0.2\,$Hz. $a_1$, $a_2$ and $a_3$ are plotted in the Fig.~\ref{modespa} (a), (b) and (c). With this definition $a_i$ is the equivalent of the amplitude of the component of the wave-field, $a_i=\frac{1}{2}\,|B_i|$. As the spectrum is a quadratic operation, we observe also a short spatial modulation at twice the wavenumber $k_i$ {due to the reflections on the tank boundary}. In (a) and (b), the mother waves decay strongly {with the distance to the  wavemakers, mainly due to the viscous dissipation}. In (c), the daughter wave at the frequency $f_3$ grows in the direction provided by $\mathbf{k_1}+\mathbf{k_2}$ in the zone of crossing of the two mother waves and then seems to be affected by the interference with reflected waves in the top of the figure. {The daughter wave amplitude is significant essentially inside the interaction zone.}

\begin{figure}[h!]
\centering
\includegraphics[width=.32\textwidth]{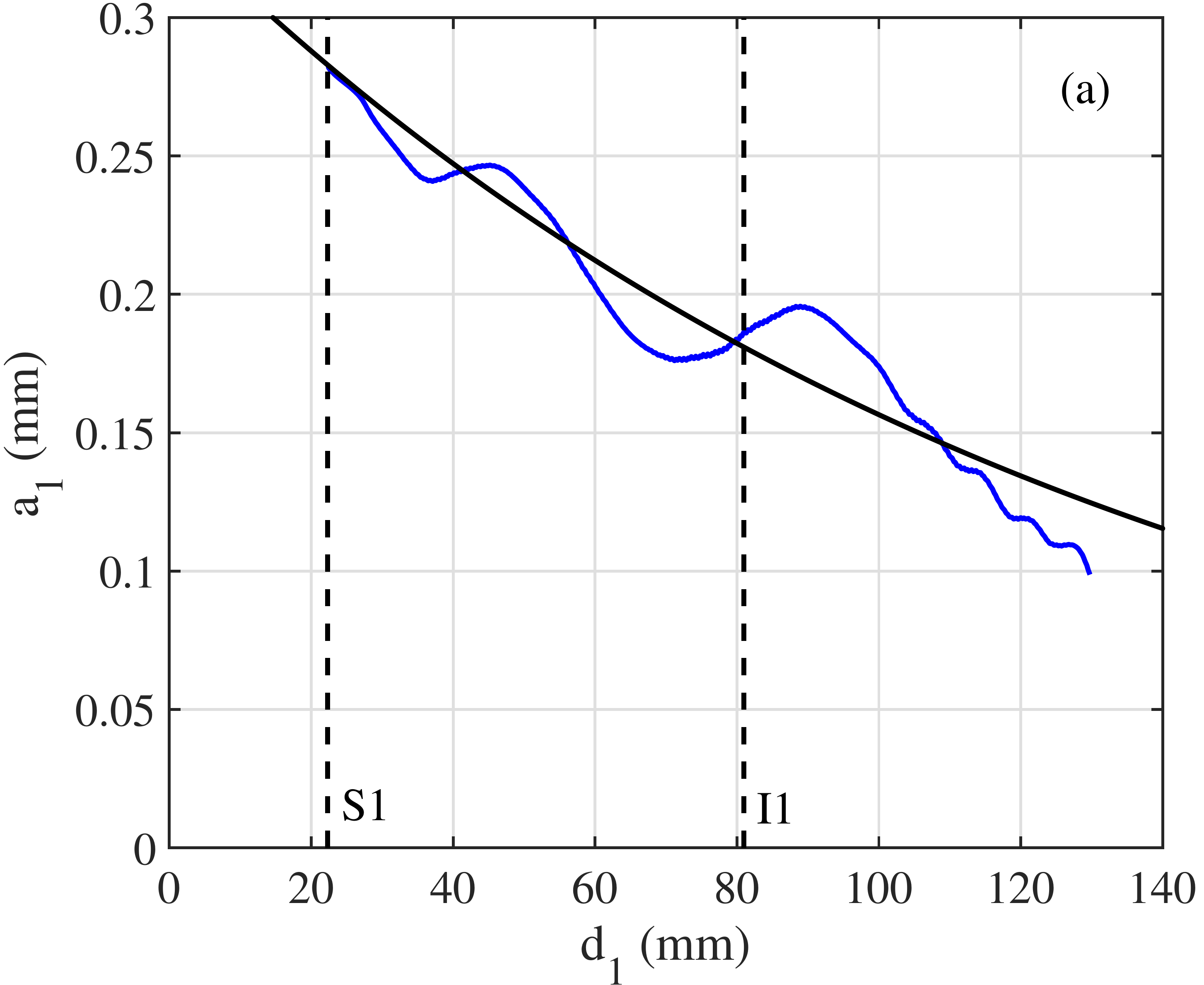}
\includegraphics[width=.32\textwidth]{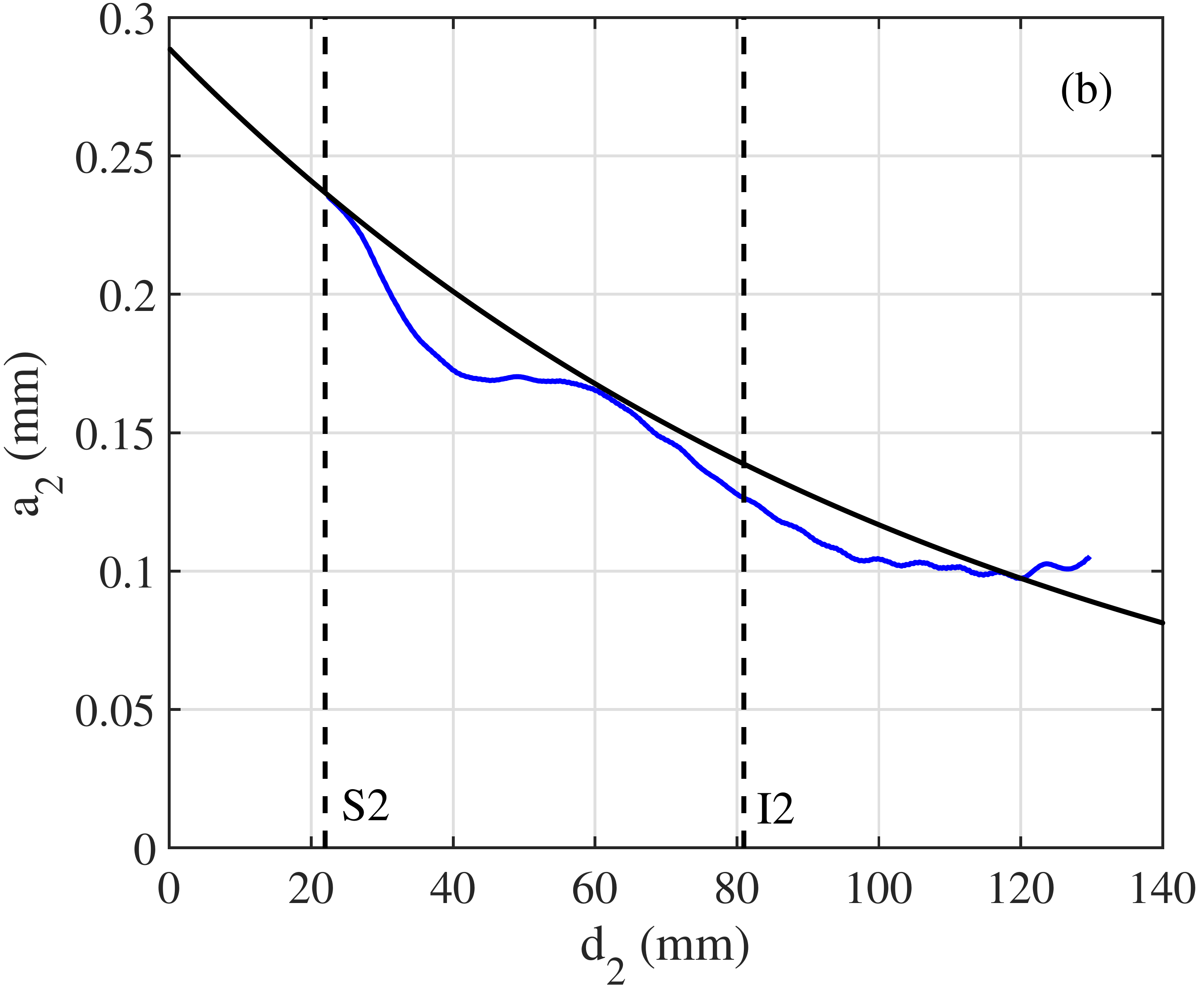}
\includegraphics[width=.32\textwidth]{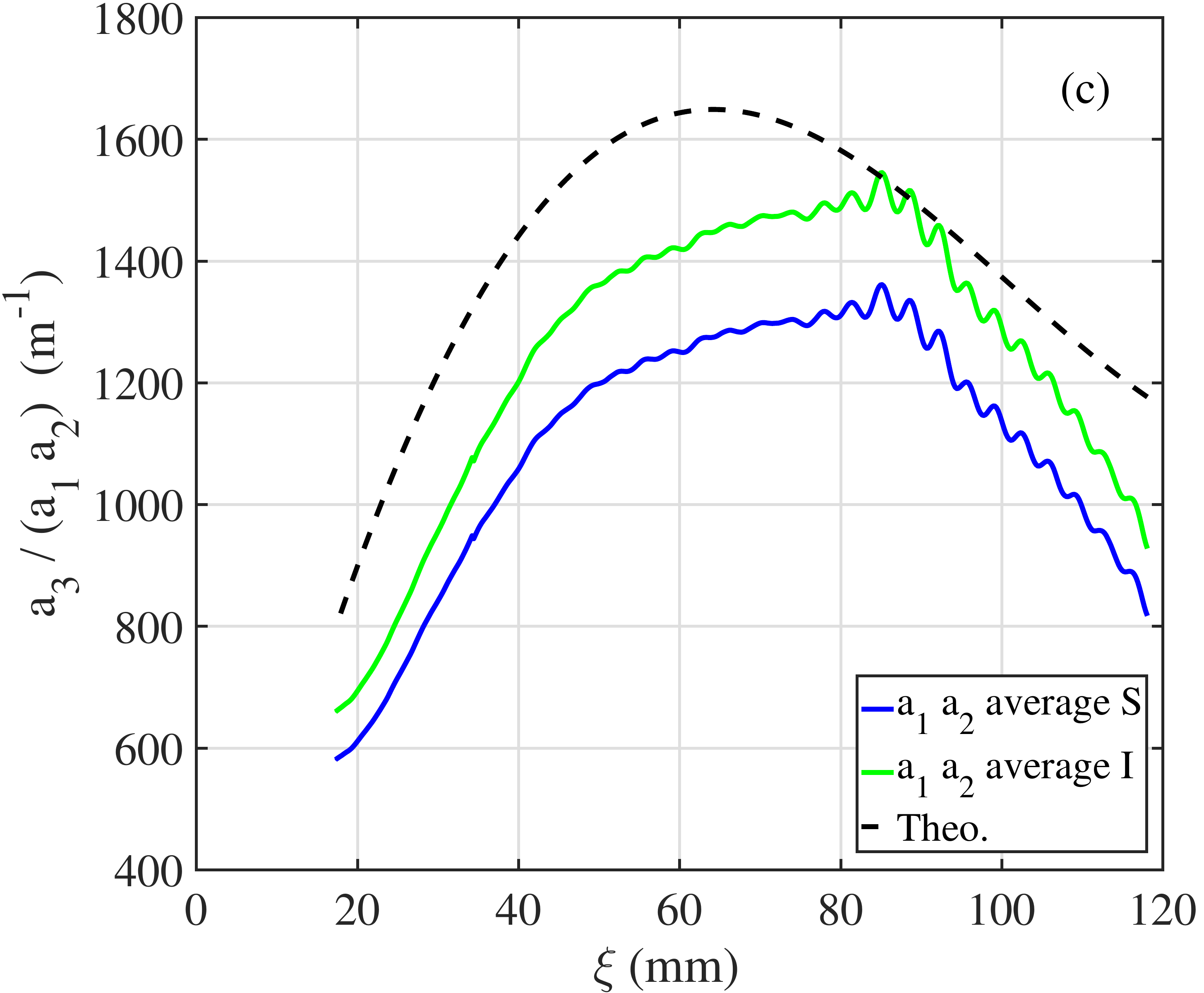}
\caption{Spatial evolution of the mother waves {$a_1$} at  $f_1$ (a) and {$a_2$} at $f_2$ (b) along their corresponding propagation direction {(black lines in Fig.~\ref{modespa} (a) and (b)) as a function of the distance to the wavemaker $d_i$. The mother waves decay strongly with $d_i$ due to the viscous dissipation. The corresponding decrease $a_i\sim \e^{(-\delta_i\,d_i/v_{g\,i})}$ is indicated with a black line. $\mathrm{S_i} $ is the position from which a part of the wave beam enters in the interaction zone (blue lines in Fig.~\ref{schema3ondes} (b)). $\mathrm{I_i} $ is the position of entrance of the mother wave in the interaction zone (see Fig.~\ref{modespa})} (c) Spatial evolution of the daughter wave amplitude {$a_3$} rescaled by the product of the average amplitude of the mother waves $\langle a_1 \rangle_{d_1} \langle a_2 \rangle_{d_2}$ as a function of $\xi$ the distance to the beginning of the interaction zone (point $O$ in Fig.~\ref{schema3ondes}) for $\alpha_{1\,2}=40\,^\circ$. {Two procedures of averaging the amplitude of the mother waves have been tested. In blue (dark gray) $\langle a_i \rangle_{d_i}$ is the average of $a_i$ from (a) or (b) in the domain $d_i > \mathrm{S_i} $. In green (light gray) the average is performed inside the interaction zone $d_i > \mathrm{I_i} $.} The theoretical solution in dashed line is given by the modulus of Eq.~(\ref{eta3full}).}
\label{modes123}
\end{figure}

Now, to compare with our model, we study the spatial evolution of the components of the triad in their respective propagation direction. For the mother waves, we consider the line starting from the middle of the corresponding wavemaker and perpendicular to it (black dashed lines in Figs.~\ref{modespa} (a) and (b)). The amplitude $a_i$ averaged on a circle of radius $5\,$mm is attributed to each point of this line, in order to evaluate the dependency of $a_i$ {as a function of the distance to the wavemaker $d_i$. The same procedure is performed to project the amplitude $a_3$ on the line $O_\xi$ (black line in Fig.~\ref{modespa} (c))}.
In Fig.~\ref{modes123} (a) (resp. (b)), the amplitude of the mother wave $a_1$ (resp. $a_2$) is displayed as a function of the distance $d_i$ to the wavemaker $1$ (resp. $2$). We observe a significant decay of these components due to the viscous dissipation. {Their amplitude declines are compatible indeed with the attenuation lengths $\delta_i/v_{g\,i}$ predicted by the inextensible film model. The decay caused by the pumping by the daughter wave seems to be of smaller importance.} Thus, the amplitudes of the mother waves are not constant, in contradiction with the hypothesis of our model. The typical  steepness $k_i\,a_i$ is of order $0.1$  at the entrance in the interaction zone and is of order $0.07$ for $d_i \approx 80\,$mm.\\

Then, the spatial behavior of the daughter wave  $a_3$ is expressed in a system of {axes}, where the origin is the beginning of the interaction zone (the crossing region between the two mother waves) similarly to our model. $\xi$ is the distance to this origin $O$ in the direction of $\mathbf{k_p}$.

To compare the measurements with the theory, the amplitude of the daughter wave must be rescaled by the product of the amplitudes of the mother waves. However, we observe that these last amplitudes are not constant in space mainly due to the viscous dissipation and to the decay with the distance to the corresponding wave maker. What is the relevant mother wave amplitude to discuss our measurements ? We can use the maximal wave amplitudes at the entrance in the interaction zone, which leads to an overestimation of the mother wave amplitudes or the average wave amplitude on the interaction zone, which leads to an underestimation.  We adopt this last approach in the following, which seems to us more consistent regarding the experimental variability. {We define the average value of the mother wave amplitude $\langle a_i \rangle_{d_i}$ as the spatial average of $a_i$ along the distance $d_i$ (see Fig.~\ref{modes123} (a) and (b)). In Fig.~\ref{modes123} (a-b), the mother wave penetrates in the interaction zone when $d_i$ is larger than the coordinate of point $\mathrm{I_i}$ defined as the first intersection of the black line with the magenta lines in Fig.~\ref{modespa} (a-b). However, due to the width of the mother wave beams, we introduce also the point $\mathrm{S_i}$ from which a part of the beam penetrates in the interaction zone. $\mathrm{S_i}$ corresponds to the intersection of the blue lines of Fig.~\ref{schema3ondes} (b) with the line starting from the center of the corresponding wavemaker. As the daughter wave  accumulates energy during its propagation along the line $O_\xi$ from the point $O$, this last approach could appear more rigorous.} Then, in Fig.~\ref{modes123} (c), the evolution of the daughter wave amplitude $a_3$ rescaled by the product of the average mother wave amplitude $\langle a_1 \rangle_d \langle a_2 \rangle_d$  is displayed as a function of $\xi$. {The two procedures of averaging the amplitude of the mother waves have been tested with an average of $a_i$ in the domain where $d_i$ is beyond than the position of the point $\mathrm{S_i}$ ($d_i > \mathrm{S_i}$) or in the domain where $d_i$ is beyond than the position of the point $\mathrm{I_i}$ ($d_i > \mathrm{I_i}$).} We observe at short distance a growth of $a_3$, followed by a zone of nearly constant amplitude, then a decay. These curves are compared with the result of our theoretical model (dashed line) the modulus of $\eta_3(\xi)$ given by { Eq.~(\ref{eta3full})}. We use in the model a viscous decay rate for the daughter wave $\delta_3= \sqrt{2\,\nu\,\omega}\,k_i/4 =4.83\,$s$^{-1}$ as the viscosity of the solution of Intralipids is $\nu=1.24 \times 10^{-6}$\,m$^{2}$ s$^{-1}$. The theory and the experiment have the same order of magnitude and present qualitatively the same behavior. {We note that by averaging of the mother waves in the domain $d_i > \mathrm{I_i}$ the rescaled daughter wave is closer from the theoretical estimate, then we use this averaging procedure in the following. We interpret the decay of the daughter wave for $\xi>80\,$mm as a part of an oscillation of the daughter wave amplitude predicted theoretically with a wavelength of order $2\pi/|k_p -  k_3|$}. This study  for the angle $\alpha_{1\,2}=40^\circ$ demonstrates that a daughter wave is well created by a forced interaction between the mother waves in the interaction zone. Our simplified model describes qualitatively the experimental data, which appears satisfying in front of the hypotheses made, which are not well verified experimentally: constant mother wave amplitude, absence of reflected waves and very weak nonlinearity. \\

\begin{figure}[h!]
\centering
\includegraphics[width=15cm]{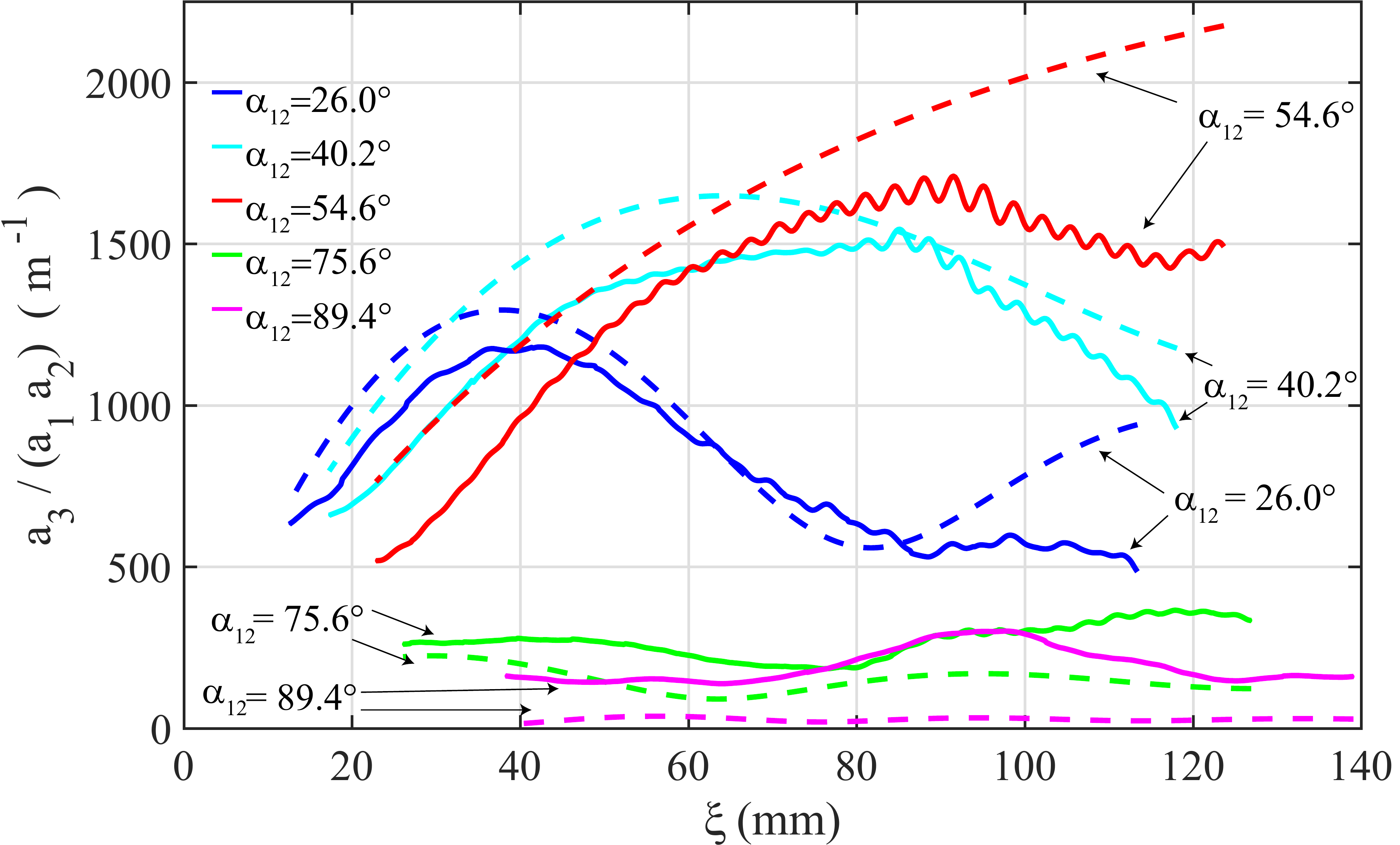}
\caption{Daughter wave amplitude $a_3$ normalized by {the product of the average amplitude of the mother waves inside the interaction zone} $\langle a_1 \rangle_{d_1}  \, \langle a_2 \rangle_{d_2}$ as a function of the distance $\xi$ at a given time $t_1$, for various values of the angle $\alpha_{1\,2}$ between the two mother waves. Continuous line experiment. Dashed lines model from the modulus of Eq.~(\ref{eta3full}).}
\label{eta3vsxmanip}
\end{figure}

\subsection{{Variation of the daughter wave amplitude with the angle $\alpha_{1\,2}$}}

We extend now the study to the other values of the angles in order to evidence the expected resonant behavior for $\alpha_{1\,2} \approx \alpha_{1\,2\,r}$, always for the triad  $f_1=15$\,Hz, $f_2=18$\,Hz and $f_3=33$\,Hz. For a constant amplitude of excitation of the mother waves, the angle between the two wave makers is varied and the same procedure is applied. Qualitatively, a same behavior is observed, for an angle between the two mother waves $\alpha_{1\,2}$ differing from the resonant angle $\alpha_{1\,2\,r} \approx 54\,^\circ$. A daughter wave of significant amplitude is detected at the frequency $f_3=f_1+f_2$  and propagates in the direction $\mathbf{k_p}=\mathbf{k_1}+\mathbf{k_2}$. The spatial behavior of the rescaled daughter wave amplitude is plotted in Fig.~\ref{eta3vsxmanip} as a function of $\xi$  for few values of  $\alpha_{1\,2}$. The measurements are compared with the predictions of the models (dashed line) for the corresponding values of experimental parameters and the same range of variation of $\xi$. After an initial growth, we observe a spatial modulation of the daughter wave, with a smaller spatial period when $\alpha_{1\,2}$ differs notably from the resonant angle. For the $\alpha_{1\,2}=26.0^\circ$, $\alpha_{1\,2}=40.2^\circ$ and $\alpha_{1\,2}=54.6^\circ$ the experimental curves are below the theoretical prediction and the contrary for $\alpha_{1\,2}=75.6^\circ$ and $\alpha_{1\,2}=89.4^\circ$. We note that the response of the free surface is larger than theoretically expected for $\alpha_{1\,2} \approx 90^\circ$. The finite size of the container and the induced reflections limit also the comparison to the first part of the tank $\xi < 100\,$mm. The qualitative behavior remains nevertheless acceptable.\\

\begin{figure}[h!]
\centering
\includegraphics[height=.4\textwidth]{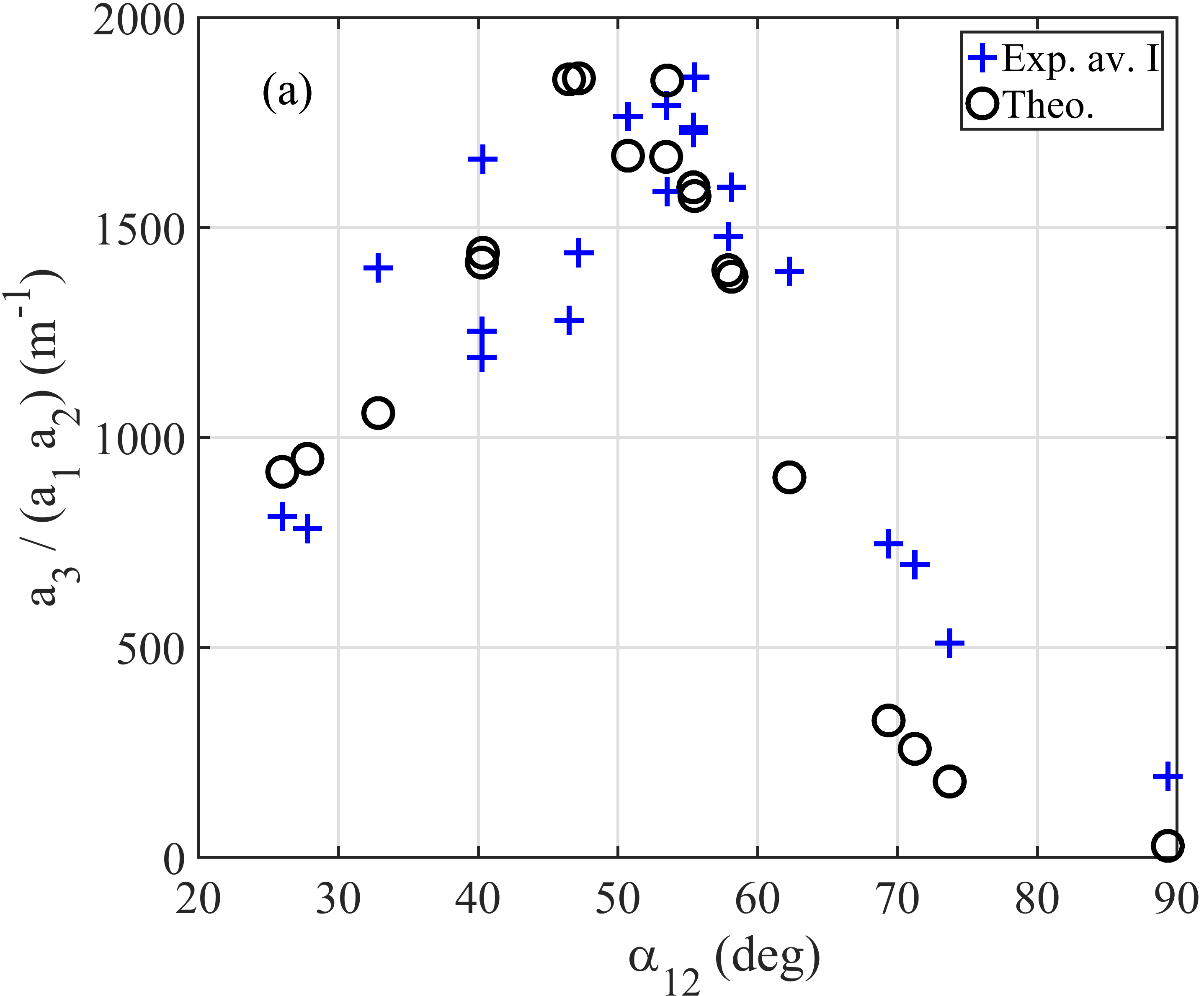}\hfill
\includegraphics[height=.4\textwidth]{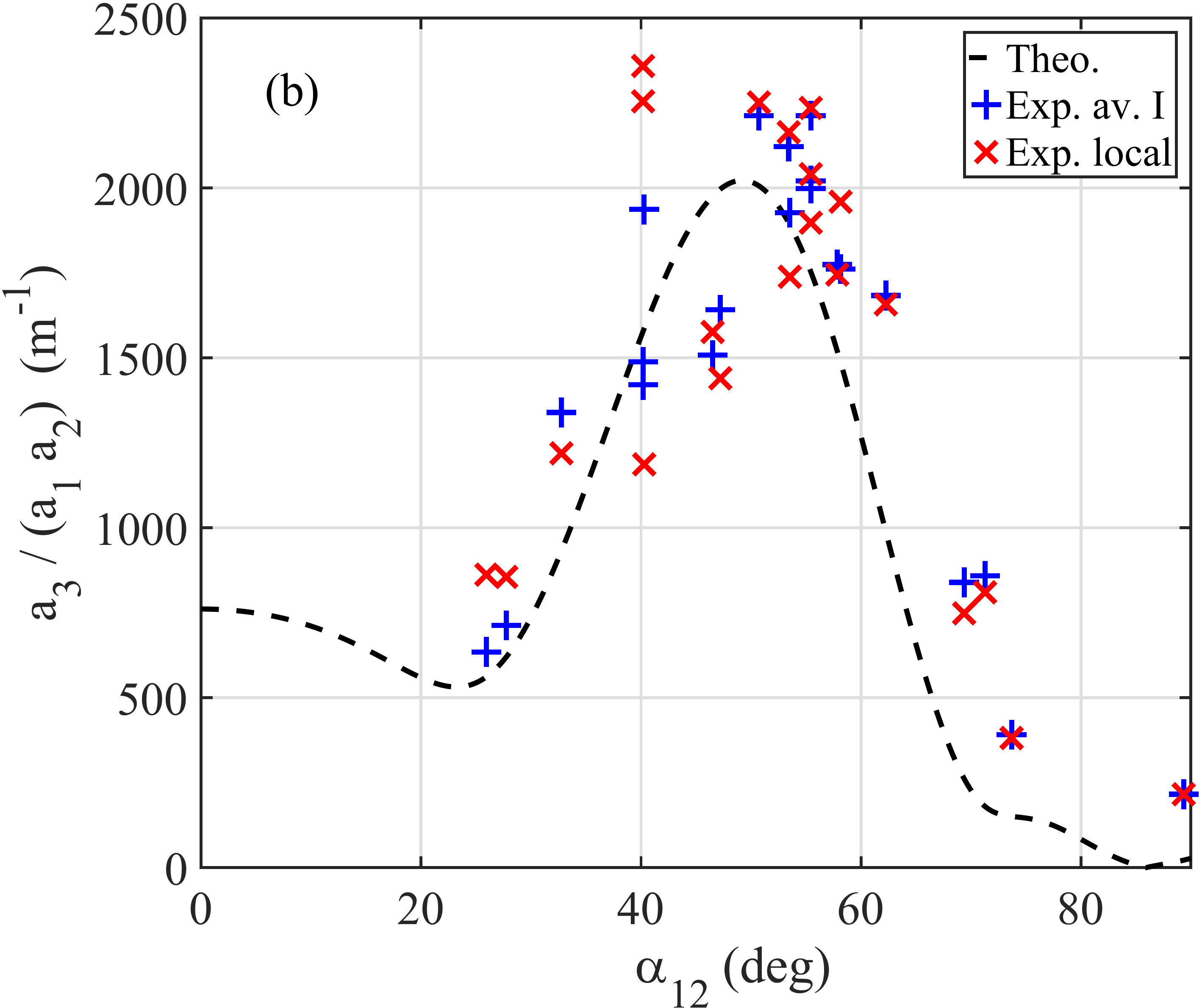}
\caption{(a) For the triad $f_1=15$, $f_2=18$ and $f_3=33$\,Hz, average value of the daughter wave amplitude $\langle a_3  \rangle_{\xi}$ rescaled by the product of the average amplitude of the mother waves inside the interaction zone $\langle a_1 \rangle_{d_1} \langle a_2 \rangle_{d_2}$ as a function of  the angle between the mother waves $\alpha_{1\,2}$. The experimental points (blue $+$) are compared with the results of the model (black circles). (b)  To take into account the spatial variation of $a_3(\xi)$ the local value of $a_3$ for $\xi=80\,$mm $a_3(\xi=80\,\mathrm{mm})$ is depicted as a function  of  $\alpha_{1\,2}$.  Blue $+$ rescaling by the average amplitude of the mother waves  inside the interaction zone $\langle a_1 \rangle_{d_1} \langle a_2 \rangle_{d_2}$. Red $\times$ rescaling by the local amplitude of the mother waves $ a_1 (\xi=80\,\mathrm{mm})\, a_2 (\xi=80\,\mathrm{mm}) $. {The typical variation between the points is about $20\%$.} The prediction of the model for $\xi=80\,$mm is plotted in dashed black line. } 
\label{testreso1518}
\end{figure}

Then, to compare the amplitude of the daughter wave for all the tested angle values, its averaged value $\langle a_3 \rangle_\xi$ is depicted in Fig.~\ref{testreso1518} (a) and compared to the theoretical prediction. {When the angle $\alpha_{12}$ is changed, the positions of the wavemakers inside the tank are also modified. Therefore, for some runs sharing a same angle value the distance between the wavemakers and so the geometry of the interaction zone can differ. For each experimental point, the theoretical estimation displayed in Fig.~\ref{testreso1518} (a) is computed for the actual spatial configuration.} The difference between the model and the experiment is of order $25\%$. 
In order to take into account the changes of the spatial configurations of wavemakers, the value of the daughter wave is plotted now in Fig.~\ref{testreso1518} (b) for a particular point at $\xi=80\,$mm, roughly in the middle of the interaction zone. For this distance, the theoretical prediction of the model is the dashed black curve which presents a maximum close to the  resonant angle $\alpha_{1\,2\,r}=54^\circ$. Two procedures of rescaling by the mother waves are tried. In the first, $a_3(\xi=80\,\mathrm{mm})$ is divided by the average values of the mother waves inside the interaction zone $\langle a_1 \rangle_{d_1} \langle a_2 \rangle_{d_2}$. This approach is closer to the hypotheses of the model, which supposes constant amplitude of mother waves. In the second,  $a_3(\xi=80\,\mathrm{mm})$ is divided by the product of the local values of the mother wave amplitudes $ a_1 (\xi=80\,\mathrm{mm})\, a_2 (\xi=80\,\mathrm{mm}) $, which corresponds to the data processing performed with a local probe in our previous work~\cite{Haudin2016}. This last method presents a larger dispersion of the experimental points, mainly due to the reflections of the mother waves. We note in all case a quite strong variability of experimental points from different experiments. The reflections at boundaries with a contact line hysteresis~\cite{Michel2016} and  changes of the wavemaker configuration  can modify the local amplitude of the free surface due to a variable amplitude of the reflected wave between the runs. By taking the averaged value of mother waves, the experimental points appear more gathered. 

{Given the uncertainty of some experimental parameters (the surface tension $\gamma$, the decay rate $\delta_3$) and the experimental variability the agreement between the model and the measurements remains acceptable by taking the average mother wave amplitude inside the interaction zone or the local mother wave amplitude. This simple model which does not take into account the decay of the mother waves, explains however qualitatively the main features of the daughter wave generated by a forced three-wave interaction.}\\

\begin{figure}[h!]
\centering
\includegraphics[height=.4\textwidth]{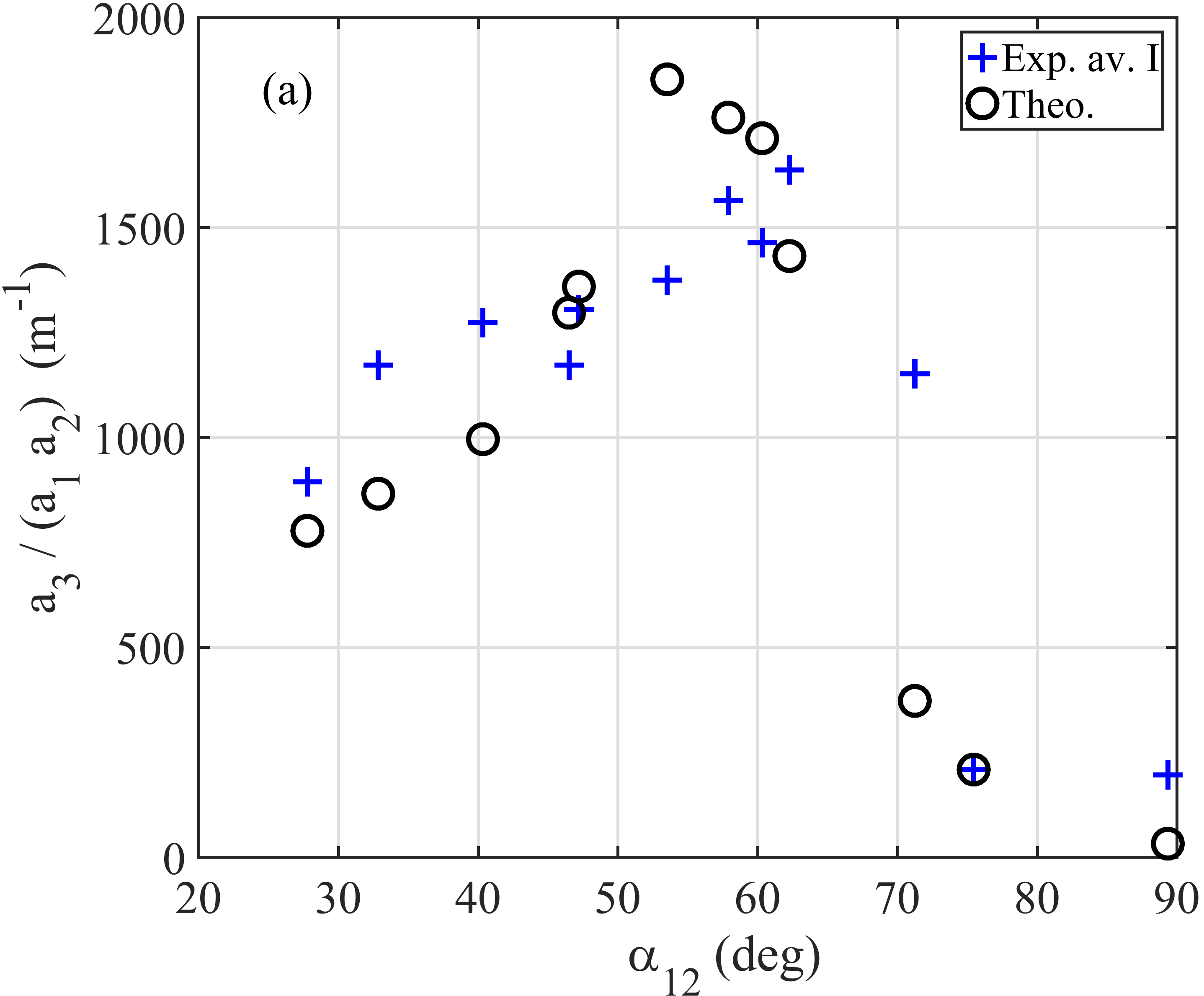} \hfill
\includegraphics[height=.4\textwidth]{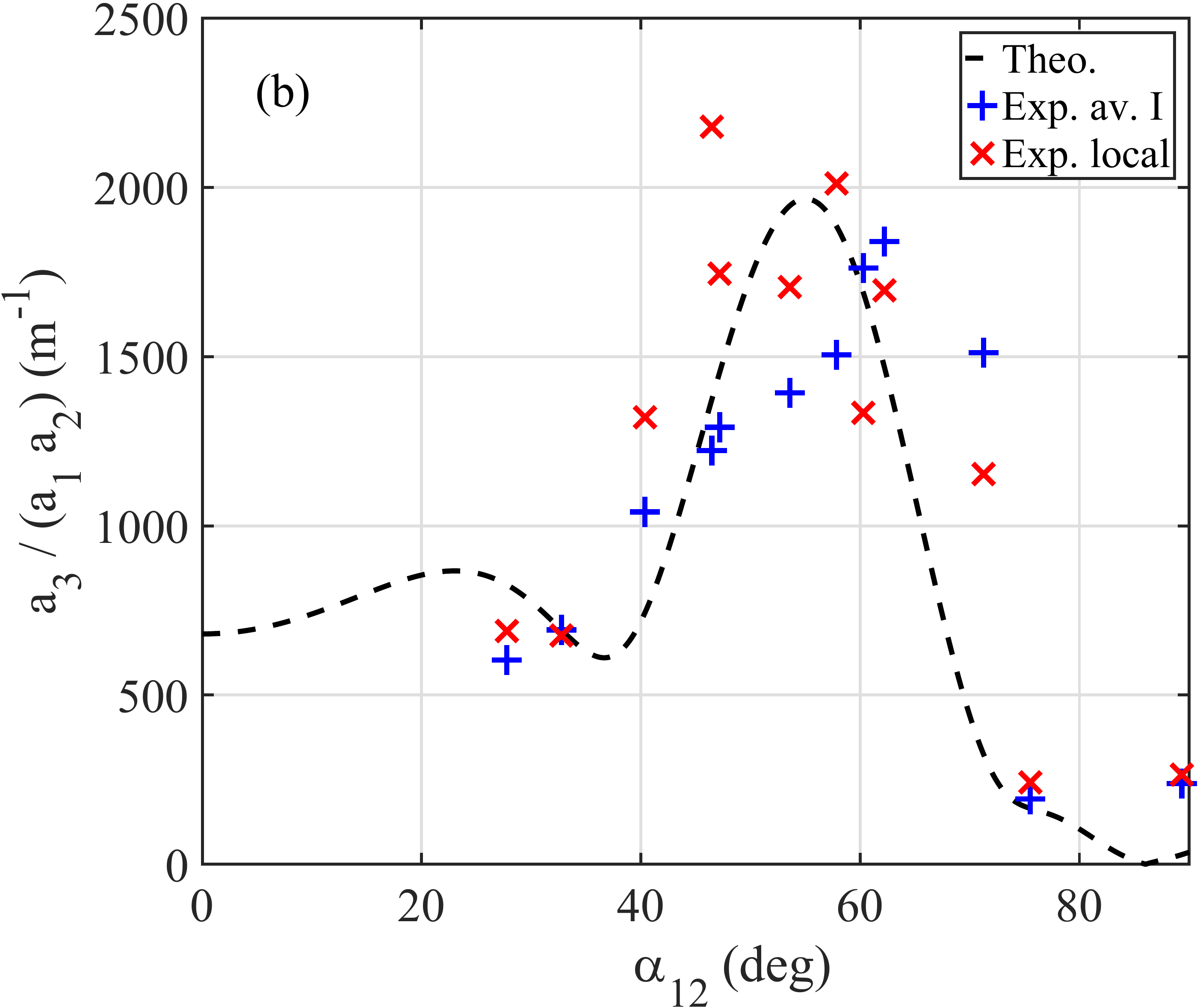}
\caption{(a) For the triad $f_1=16$, $f_2=23$ and $f_3=39$\,Hz, average value of the daughter wave amplitude $\langle a_3  \rangle_{\xi}$ rescaled by the product of the average amplitude of the mother waves $\langle a_1 \rangle_{d_1} \langle a_2 \rangle_{d_2}$ as a function of  the angle between the mother waves $\alpha_{1\,2}$. The experimental points (blue $+$) are compared with the results of the model (black circles). (b)  To take into account the spatial variation of $a_3(\xi)$ the local value of $a_3$ for $\xi=80\,$mm $a_3(\xi=80\,\mathrm{mm})$ is depicted as a function  of  $\alpha_{1\,2}$.  Blue $+$ rescaling by the average amplitude of the mother waves $\langle a_1 \rangle_{d_1} \langle a_2 \rangle_{d_2}$. Red $\times$ rescaling by the local amplitudes of the mother waves $ a_1 (\xi=80\,\mathrm{mm})\, a_2 (\xi=80\,\mathrm{mm}) $. {The typical variation between the points is about $20\%$.} The prediction of the model for $\xi=80\,$mm is plotted in dashed black line. }
\label{testreso1623}
\end{figure}

Another smaller set of experiments has been tested for the triad $f_1=16$\,Hz, $f_2=23$\,Hz and $f_3=39$\,Hz, with the same experimental protocol. The resonant angle is then $\alpha_{1\,2\,r}=59^\circ$. The results and the observations are very similar for both triads with the same qualitative behavior. To compare with the model predictions, the average daughter wave amplitude $\langle a_3  \rangle_{\xi}$ in Fig.~\ref{testreso1623} (a) and the local daughter wave amplitude $a_3(\xi=80\,\mathrm{mm})$ in Fig.~\ref{testreso1623} (b) are plotted. Again, we find a qualitative agreement.

\section{Conclusion and perspectives}

In this article, we have first demonstrated experimentally the generation of capillary waves by a three-wave resonant mechanism, in conditions where they are unexpected because these generated waves do not follow exactly the linear dispersion relation. These observations can be understood by the concept of forced three-wave resonant interactions. The two mother waves interact as a product, which induces an excitation of the free surface at the sum frequency and the sum wavenumber. These sums match the dispersion relation only for a specific angle between the mother waves $\alpha_{1\,2\,r}$. As the dispersion relation characterizes the propagation of waves in absence of forcing, the system response is maximal close to $\alpha_{1\,2\,r}$, {similarly to the case of a classic forced oscillator for which an excitation close to its eigen-mode induces a resonance phenomenon.} Similarly, the bandwidth of the resonance increases also with the dissipation rate of the daughter wave. A weakly nonlinear model of the response of the free surface provides an analytic expression of the daughter wave amplitude as a function of its propagation distance, but in the limit of constant amplitude of mother waves. The concept of forced resonant wave interactions is very analogous to the non or quasi resonant interactions, where the dispersion relation is followed but one of the resonance condition is not verified. Then, our theoretical results are confronted to another set of experiments where the angle $\alpha_{1\,2}$ between the mother waves has been varied. Despite the viscous decay of the mother waves, the finite width of the wave beams and the reflections on the boundaries of the small tank, the experimental behavior of the daughter wave is qualitatively well explained, validating the physical explanation of the observations by a forced resonant interaction. To improve the quantitative agreement, numerical direct simulations of this problem would be useful  and could test more deeply the concept of forced three-wave resonant interactions. The variation of the mother wave amplitudes due to the viscous decay and to the nonlinear energy pumping by the third wave could be thus specifically investigated. {According to our model, the main physical ingredient needed to observe a significant response for a forced interaction when $\alpha_{1\,2}\neq \alpha_{1\,2\,r}$ is the non-negligible wave dissipation. In this case, at short distance (\textit{i.e.} small values of $\xi$ in front of $v_{g\,3}/\delta_3$) daughter waves of largest amplitude occur indeed for $\alpha_{1\,2} < \alpha_{1\,2\,r}$ as shown in Fig.~\ref{distreso} (c).} With the finite width of mother wave beams, the zone of wave interaction is also spatially delimited. Moreover, the dissipation of the mother waves limits also the relevant size of the system to a dozen of centimeters, when the mother waves are generated in the capillary range. The measurements are then performed close to the generation area, where non-resonant responses are possible. These responses can indeed be said non-resonant in regard to the dispersion relation as explained here but also in front of the resonant conditions like the usual non-resonant or quasi-resonant wave interactions~\cite{Janssen2004,Bonnefoy2016}. Therefore in the statistical study of capillary-gravity waves in weakly nonlinear interaction, due to the viscous dissipation, the contribution of forced interactions and non-resonant interactions should be thus not neglected.

 {For laboratory generated waves, the size and the shape of the container can also play a role. }Recently, an experimental and theoretical study~\cite{Michel2019} has {indeed} demonstrated that in a confined cylindrical geometry {the large modes of gravity surface waves interact with each other by a three-wave resonant process, whereas plane gravity waves are subjected only to the four-wave resonant interaction mechanism~\cite{Janssen2004}}. Here, despite the cylindrical shape of the container, the mother waves are forced as plane waves and at frequencies significantly larger than those corresponding to the first tank eigen-modes. The dissipation is also too large to generate an eigen-mode of high order after multiple reflections~\cite{Berhanu2018}. Thus, this new interesting mechanism appears thus non relevant for our experimental configuration.\\
{The forced interaction mechanism presents also some analogies with the bound waves~\cite{Janssen2004}, which are created by the three-wave interaction of a nonlinear gravity with itself and are not following the linear dispersion relation. Recently, in the propagation of a coherent wave group, quasi-resonances generating bound waves have been  reported~\cite{Slunyaev2018}, by performing numerical simulations of the full nonlinear Euler equations without dissipation. This work constitutes another example with a different mechanism where nonlinear wave interactions lead to the generation of excitations of the free-surface not following the linear dispersion relation.}\\

{The forced three-wave resonant interaction mechanism evidenced here for capillary-gravity surface waves can be applied to other physical systems where waves interact nonlinearly like the waves in plasmas~\cite{Ritz1988,Sokolov2014}. This mechanism could occur in hydrodynamics for inertial waves~\cite{Bordes2012} and for internal waves~\cite{Joubaud2012} especially in the laboratory experiments, for which the viscous dissipation is not negligible. In our previous study of hydro-elastic waves, we reported~\cite{Deike2017} a resonant three-wave interaction of stretching waves. These waves are formally described in the weakly nonlinear limit in the same formalism than capillary waves and the present work can be used directly to this experimental situation, when the angle between the mother waves is not the resonant angle. The interaction between surface waves and a hydrodynamic flow can be also modeled at weak amplitude as a three-wave interaction~\cite{Craik}. In particular the generation of cross-waves in capillary regime may be explained by a three-wave resonant interaction between the stationary cross-wave and the oscillating flow in the vicinity of the wavemaker~\cite{Moisy2012}. Moreover, we have modeled the forced three-wave resonant interaction mechanism by the linear transfer function of the free-surface forced nonlinearly by the product of the mother waves. A similar approach where the free-surface is modeled by a linear transfer function has been also reported in a recent study of the interaction of a turbulent air flow with a water surface~\cite{PerrardArxiv} ; the turbulent fluctuations are filtered by the free-surface.}\\

Finally, we consider the case of wave turbulence where a set of random waves interact nonlinearly through three-wave interactions. This theory presupposes a scale separation  $\tau_L \ll \tau_{NL} \ll \tau_d $ between the linear time $\tau_L$ (the wave period), the nonlinear time $\tau_{NL}$ (the typical time of wave interaction decreasing with wave amplitude) and the dissipation time $\tau_d$ (the decay time of the wave due to viscosity \textit{i.e.} $\delta^{-1}$). However, in the case of capillary waves, the significant wave dissipation imposes a quite small value of $\tau_d$ of order of the second in the capillary cascade. Consequently, in experiments showing turbulent spectra, these three times are not well separated~\cite{Berhanu2018}. Moreover the characteristic system size is limited due to the finite dimensions of the tank and also due to the attenuation length of order $10\,$cm or less for capillary waves. Then, according to our model, at short distance from the generation point, one-dimensional (involving colinear vectors) forced three-wave interactions are favored, because at short distance the maximal response occurs for $\alpha=0^\circ$, as it can be seen in Fig.~\ref{distreso} (c). The predominance of one-dimensional three-wave interactions have been reported in experiments studying wave turbulence for gravity waves close to the gravity-capillary crossover~\cite{Aubourg2015,Aubourg2016}.  
The occurrence of forced three-wave resonant interactions and of quasi-resonant interactions are thus not excluded for gravity-capillary wave turbulence experiments, for which the dissipation is not negligible. As the wave turbulence theory incorporates only purely resonant interactions, a quantitative deviation is thus observed. Future theoretical works including the dissipation directly in the computation of interaction coefficients and addressing the wave turbulence regimes in presence of non-negligible dissipation will be thus worthwhile~{to understand the dynamics of disordered ripples in the nature}.

\begin{acknowledgments}
We thank Luc Deike for scientific discussions and Alexandre Lantheaume for technical assistance. This work was funded by ANR-12-BS04-0005 Turbulon. A. Cazaubiel and E. Falcon acknowledge the funding by ANR-17-CE30-0004 Dysturb.
\end{acknowledgments}

\appendix
\section{Weakly nonlinear response of a free surface}
\label{Modelderivation}
We aim to determine {here} the response of the liquid free surface in presence of two mother capillary-gravity waves interacting nonlinearly. {We consider the Cartesian coordinates system $O_{xyz}$, where $O_z$ is the vertical axis and the plane $z=0$ corresponds to the position the air-water interface at rest in absence of waves.} Following the approach of Case and Chiu~\cite{Case1977}, we consider the deep water limit  (bottom $z \rightarrow - \infty$), an infinite system, a potential flow (absence of vorticity except inside a thin boundary layer at the surface) and we keep only the first order in the nonlinear development, valid for weak nonlinearity. We neglect the air density above the free surface and the dynamics of the free surface is described by two scalar field: $\varphi (x,y,z)$ is the velocity potential related to the velocity field  $\mathbf{v}$ by $\mathbf{v}=\mathbf{\nabla } \varphi$ and $\eta (x,y)$ is the free surface deformation {($\varphi (x,y,z)=0$ and $\eta(x,y)=0$ at rest)}. We note that the condition of weak nonlinearity implies a small wave steepness, $|| \mathbf{\nabla}\eta || \ll 1$. Without vorticity, $\varphi$ obeys a Laplace equation in the liquid $\triangle \varphi = 0$, {in the domain $z<\eta$}.\\

At the free surface $z=\eta$,  the Bernoulli equation gives:
\begin{equation}
\dfrac{\partial \varphi}{\partial t}+\dfrac{(\mathbf{\nabla } \varphi)^2}{2}=\dfrac{\gamma}{\rho}\,\kappa (x,y,t)-g \,\eta  
\label{Bernoullisurface}
\end{equation}
where $\kappa (x,y,t)$ is the interface curvature.\\

Moreover, the kinematic boundary condition is written at the interface at $z=\eta$:
\begin{equation}\dfrac{\partial \eta}{\partial t}+\mathbf{ \nabla_r} \varphi \cdot \mathbf{ \nabla_r} \eta =  \dfrac{\partial \varphi}{\partial z}\label{kinematiccondsurface}
\end{equation}

We perform a Taylor expansion {at the free surface $z=\eta$, close to $z=0$ at first order in $\eta$}: $$\varphi(x,y,z=\eta)=\varphi(x,y,z=0)+\dfrac{\partial \varphi}{\partial z}\big{\vert}_{z=0}\,\,\eta\,$$ The curvature $\kappa$ is also assimilated to the horizontal 2D Laplacian.\\

 One obtains then the following system valid at second order in $\eta$ and $\varphi$, where nonlinear terms are gathered in the right hand side:

\begin{eqnarray}
\dfrac{\partial \varphi}{\partial t} -\dfrac{\gamma}{\rho}\, \left(\dfrac{\partial^2 \eta}{\partial x^2}+\dfrac{\partial^2 \eta}{\partial y^2} \right)+ g \,\eta & = & -\dfrac{(\mathbf{\nabla} \varphi)^2}{2} - \dfrac{\partial^2 \varphi}{\partial z \partial t}\,\eta \quad \quad \quad \qquad  \quad{z=0} \label{syst1} \\
\dfrac{\partial \eta}{\partial t} -\dfrac{\partial \varphi}{\partial z}& =  &\dfrac{\partial^2 \varphi}{\partial z^2}\,\eta - \dfrac{\partial \varphi}{\partial x}\dfrac{\partial \eta}{\partial x}-\dfrac{\partial \varphi}{\partial y}\dfrac{\partial \eta}{\partial y} \quad \quad \quad  {z=0} \label{syst2}\\
\dfrac{\partial^2 \varphi}{\partial x^2}+\dfrac{\partial^2 \varphi}{\partial y^2}+\dfrac{\partial^2 \varphi}{\partial z^2}& = &0 \quad \quad \quad  \quad \quad\qquad  \qquad  \qquad  \qquad  {-\infty < z \leq 0} \label{syst3}
\end{eqnarray}

\newpage

\section{Amplitude equation of the mode $3$ forced by the waves $1$ and $2$}
\label{Amplitudeequation3}
One supposes that the two mother waves are crossing each other with the angle $\alpha_{1\,2}$ and that their respective amplitudes $B_1$ and  $B_2$ are constant (not varying with space and with time). We search for the linear response of the free surface $\omega_3=\omega_1+\omega_2$ forced by the mother waves. In the right hand side of Eqs.~(\ref{syst1}) and (\ref{syst2}), nonlinear terms imply the product of the two mother waves and act as a forcing term on the left hand side.\\

Using the complex formalism, we consider the linear superposition of the three waves:
$$\underline{\eta}=\underline{\eta_1 }+\underline{\eta_2}+ \underline{\eta_3} \quad \quad  \mathrm{and}  \quad \quad \underline{\varphi}=\underline{\varphi_1}+\underline{\varphi_2}+\underline{\varphi_3}  $$
$$ \mathrm{with}  \quad \quad \underline{\varphi_1}=A_1\,\e^{i\,(\omega_1 t- k_{1x} x -k_{1y} y )}\,\e^{k_1 z} + \mathrm{c.c.}\, ,\quad \quad \underline{\varphi_2}=A_2\,\e^{i\,(\omega_2 t- k_{2x} x -k_{2y} y )}\,\e^{k_2 z}+ \mathrm{c.c.}\, ,$$
 
$\mathrm{c.c.}$ is the complex conjugate of the left part. With this convention, the free surface reads in the real space as $\eta(x,y,t)=\frac{1}{2}\, \mathfrak{Re} (\underline{\eta}(x,y,t)) $. 
 
 $$\underline{\eta_1}=B_1\,\e^{i\,(\omega_1 t- k_{1x} x -k_{1y} y )}+ \mathrm{c.c.},\quad \quad \quad \underline{\eta_2}=B_2\,\e^{i\,(\omega_2 t- k_{2x} x -k_{2y} y )}+ \mathrm{c.c.} \, ,$$
 $$ \mathrm{with}  \quad \quad k_1^2=k_{1x}^2+k_{1y}^2 \, , \quad \quad k_2^2=k_{2x}^2+k_{2y}^2\, , \quad  \quad  A_1=i\,\dfrac{\omega_1}{k_1}\,B_1 \quad \mathrm{and} \quad A_2=i\,\dfrac{\omega_2}{k_2}\,B_2\, .$$

Arbitrarily, we change and choose the orientation of axis such as the $x-$axis is aligned with $\mathbf{k_p}=\mathbf{k_1}+\mathbf{k_2}$ (on this appendix \textbf{B} only). The new axis $O_x$ corresponds to the axis $O_\xi$ in the main text.\\
Then, $k_p=k_{px}=k_{1x}+k_{2x}$ and $k_{py}=k_{1y}+k_{2y}=0$. 

{One supposes that the free surface is forced at the wavenumber $\mathbf{k_p}$ and at the angular frequency $\omega_3=\omega_1+\omega_2$, with $k_p \neq k_3$ defined  by the linear dispersion relation $\omega_3=\sqrt{g\,k_3+\dfrac{\gamma}{\rho}\,k_3^3}$\,. Supposing an interaction zone of infinite extension, we look for a plane wave solution propagating along the $x$-axis, then the phase of the wave is supposed to not depend on $y$. As the amplitudes of mother waves are supposed constant inside the interaction zone, the envelope of the wave 3 is supposed also not to depend on the coordinate $y$. By analogy with the classic solution for surface waves in deep water regime and with the hypothesis of slowly varying envelope, we write the mode 3 as:}
{\begin{equation}
\underline{\varphi_3}=A_3(x,z,t)\,\e^{k_p\,z}\,\e^{i(\omega_3 t - k_p x)
\quad \quad \underline{\eta_3}=B_3(x,t)\,\e^{i(\omega_3 t - k_p x)}} \label{envelopes}
\end{equation}}
{$A_3(x,y,z,t)$ and $B_3(x,y,t)$ are the amplitudes or envelopes respectively of the velocity potential and of the free-surface deformation for the forced mode oscillating at $\omega_3$.}\\

Then, from the system of Eqs~\ref{syst1}, \ref{syst2} and \ref{syst3}, we get the evolution equations of the mode of angular frequency $\omega_3$:

\begin{eqnarray}
\dfrac{\partial \underline{\varphi_3}}{\partial t} -\dfrac{\gamma}{\rho}\, \dfrac{\partial^2 \underline{\eta_3}}{\partial x^2}+ g \,\underline{\eta_3} & =& \left[ \dfrac{\omega_1 \,\omega_2}{k_1\,k_2 } (k_1\,k_2 - \mathbf{k_1} \cdot \mathbf{k_2}) + \omega_1^2+\omega_2^2\right]\,B_1\,B_2\,\e^{i(\omega_3 t - k_p x)}  \label{Eqmode3a}\\
\dfrac{\partial \underline{\eta_3}}{\partial t} -\dfrac{\partial\underline{\varphi_3}}{\partial z}& =  &\, i\,\left[ \omega_1 \left(k_1+\dfrac{\mathbf{k_1} \cdot \mathbf{k_2}}{k_1} \right)+\omega_2 \left(k_2+\dfrac{\mathbf{k_1} \cdot \mathbf{k_2}}{k_2}\right) \right]\, B_1\,B_2\,\e^{i(\omega_3 t - k_p x)}\label{Eqmode3b} \\
\dfrac{\partial^2 \underline{\varphi_3}}{\partial x^2}+\dfrac{\partial^2 \underline{\varphi_3}}{\partial z^2}& = &0 \label{Eqlapphi}
\end{eqnarray}
{One can verify that the expressions of the mode 3 (Eqs.~\ref{envelopes}) are solution of the Laplace equation Eq.~(\ref{Eqlapphi}). }\\

To simplify the notations, one sets:
\begin{eqnarray}
   T_A &=& \dfrac{\omega_1 \,\omega_2}{k_1\,k_2 } (k_1\,k_2 - \mathbf{k_1} \cdot \mathbf{k_2}) + \omega_1^2+\omega_2^2 \label{coeffTA}\\ 
  \mathrm{and} \,\quad T_B &=& \omega_1 \left(k_1+\dfrac{\mathbf{k_1} \cdot \mathbf{k_2}}{k_1} \right)+\omega_2 \left(k_2+\dfrac{\mathbf{k_1} \cdot \mathbf{k_2}}{k_2}\right).
\end{eqnarray}

\noindent Spatial and temporal dependencies of amplitudes $A_3$ and $B_3$ are considered, with the following boundary conditions:  $A_3(z) \rightarrow 0$ when $z\rightarrow -\infty$, $A_3(x=0)=0$ and {$B_3(x=0)=0$}. The daughter wave domain of existence starts indeed from the  point $x=0$, beginning of the interaction zone and propagating along the growing values of $x$. As we look for an amplitude equation slowly varying in time and in space, the second derivative of $A$ and $B$ are neglected. From Eq.~(\ref{Eqlapphi}), we have:
$$\left(-2 i\,k_p \dfrac{\partial A_3}{\partial x}-k_p^2\, A_3 \right)\e^{k_p\,z}\,\e^{i(\omega_3 t - k_p x)}+ \left(2 \,k_p \dfrac{\partial A_3}{\partial z}+k_p^2\, A_3 \right)\e^{k_p\,z}\,\e^{i(\omega_3 t - k_p x)}=0 $$
We have thus $ \dfrac{\partial A_3}{\partial z}=i \dfrac{\partial A_3}{\partial x}$, a similar relation being used also in \cite{Case1977}.\\
Reporting {this last relation} in Eqs.~\ref{Eqmode3a} and \ref{Eqmode3b} evaluated in  $z=0$, one obtains:
\begin{eqnarray}
 i \omega_3 A_3+\dfrac{\partial A_3}{\partial t} +\dfrac{\gamma}{\rho} \left(2 i k_p \dfrac{\partial B_3}{\partial x} +k_p^2 B_3 \right) + g B_3 & =& T_A\,B_1\,B_2  \label{Eq7}\\  i \omega_3 B_3+\dfrac{\partial B_3}{\partial t} -k_p\,A_3 -   \dfrac{\partial A_3}{\partial z}&=&i T_B \,B_1\,B_2 
\end{eqnarray}
Knowing that the second member is supposed constant, by time deriving the last equation and always neglecting the second order derivative, one finds:
 $i \omega_3  \dfrac{\partial B_3}{\partial t}= k_p \dfrac{\partial A_3}{\partial t}  $.\\
Similarly by deriving with respect to $x$:
  $i \omega_3  \dfrac{\partial B_3}{\partial x}= k_p \dfrac{\partial A_3}{\partial x}  $.\\
Moreover, $A_3= \dfrac{i \omega_3 }{ k_p} B_3+\dfrac{1}{k_p}\,\dfrac{\partial B_3}{\partial t}+\dfrac{i}{k_p} \dfrac{\partial A_3}{\partial x} - \dfrac{i}{k_p}T_B \,B_1\,B_2 $. \\
By reporting in Eq.~(\ref{Eq7}), we obtain:
$$B_3 \left(\dfrac{-\omega_3^2}{k_p}+\dfrac{\gamma}{\rho} k_p^2+g \right)+\dfrac{\partial B_3}{\partial t}\left( \dfrac{2i\,\omega_3}{k_p}\right)+\dfrac{\partial B_3}{\partial x}\left(\dfrac{i\,\omega_3^2}{k_p^2}+\dfrac{2 i \gamma\,k_p}{\rho} \right)=T_A\,B_1\,B_2-\dfrac{\omega_3}{k_p} T_B\,B_1\,B_2  $$

Finally, after simplifications, one finds
\begin{equation}
 \dfrac{\partial B_3}{\partial t}+\dfrac{1}{2}\left(\dfrac{\omega_3}{k_p}+\dfrac{2 \gamma\,k_p^2}{\rho\,\omega_3} \right)\dfrac{\partial B_3}{\partial x}-i \left(\dfrac{\omega_p^2-\omega_3^2}{2\,\omega_3}\right)B_3=\dfrac{-i}{2}\,\left( \dfrac{k_p}{\omega_3} T_A-T_B\right) \,B_1\,B_2
 \label{B3ap}
\end{equation}
Here $\omega_p=\sqrt{g\,k_p+\dfrac{\gamma}{\rho}\,k_p^3}$\, is the angular frequency given by the linear dispersion relation for the wavenumber $k_p$. 

Therefore, as in \cite{Case1977}, we find an amplitude equation for the temporal variation of the daughter wave amplitude $B_3$. We note that the temporal variation is balanced with a spatial advection term, where the gradient of $B_3$ is multiplied by a propagation term interpreted as the velocity of the envelope:

\begin{equation}
v_p = \dfrac{1}{2}\left(\dfrac{\omega_3}{k_p} +\dfrac{2 \gamma\,k_p^2}{\rho\,\omega_3} \right) 
\label{vp}
\end{equation}
 We can indeed check that for $k_p=k_3$, $v_p=v_{g,3}$, the group velocity of the wave 3 according to the linear dispersion relation:
$$v_{g,3}=\dfrac{\partial \omega_3}{\partial k_3}= \dfrac{g\,k_3+3\dfrac{\gamma}{\rho}\,k_3^2}{2\,k_3\,\omega_3}$$\\
Moreover, the factor of $B_3$ is a pure imaginary number and induces an oscillation if $\omega_p \neq \omega_3$ and so if $k_p \neq k_3$. This kind of modulation of the envelope is typical of non-resonant interactions~\cite{Boydbook,Bonnefoy2017}.\\
The last ingredient consists to include the dissipation as a perturbation like in experimental works about three-wave interactions for gravity-capillary waves~\cite{Haudin2016,Henderson1987_1,McGoldrick1970}, by adding a term $\delta_3\,B3$, with $\delta_3$ the decay rate at the frequency $f_3$. We obtain then the amplitude equation labeled Eq.~(\ref{EqevolB3}) in the main text:
\begin{equation}
 \dfrac{\partial B_3}{\partial t}+v_p\,\dfrac{\partial B_3}{\partial x}+ \left[ \delta_3 - i \left(\dfrac{\omega_p^2-\omega_3^2}{2\,\omega_3}\right) \right] B_3=\dfrac{-i}{2}\,\left( \dfrac{k_p}{\omega_3} T_A-T_B\right) \,B_1\,B_2
 \label{EqevolB3b}
\end{equation}

\bibliography{threewavesart}

\end{document}